# Why do we need to complement the European Union Regional Innovation Scoreboard with an artificial intelligence tool for what-if policy analysis?

Vincenzo Lanzetta,[1,‡] Cristina Ponsiglione,[2,‡]

1 Information Technology and Electrical Engineering Department, University of Naples Federico II
2 Department of Industrial Engineering, University of Naples Federico II
‡ Joint first authors

To whom correspondence should be addressed; E-mails: {vincenzo.lanzetta, cristina.ponsiglione}@unina.it

## Abstract

The European Union Regional Innovation Scoreboard (EURIS) is currently and broadly used for the definition of regional innovation policies by European policymakers; it is a regional innovation measuring tool for the analysis of each specific innovation indicator, from which it is possible to analyze the overtime evolution of each regional innovation indicator; according to the importance of the European Union Regional Innovation Scoreboard for innovation policy purposes, we state that European regional policymakers need integrative and synergistic methodological tools, with respect to the EURIS one, for innovation policy purposes. We state the need to integrate the current methodology of the European Regional Innovation Scoreboard with a Factorial K-means (FKM) tool for grouping purposes, and with a neural network (NN) tool for performing what-if policy analyses. Experimental results suggested that our proposed grouping/labeling methodologies are able to develop more compact groups, resulting in regions having better similarities, than the ones developed by the EURIS methodology. Experimental simulations, within the framework of our proposed what-if tool, highlight the potential usefulness of our methodological tool (neural network-based) for the understanding of the potential effectiveness of each possible and specific regional innovation policy to be implemented. We claim that our proposed FKM-NN tool could be used, by regional innovation policymakers, as a very effective synergistic instrument of the European Union Regional Innovation Scoreboard.

# 1. Introduction

Innovation can be considered an engine of the territorial development and a key driving force for the long-term economic growth; indeed, innovative processes are able to promote competitive efficiency (Tödtling et al., 2022) and to improve the regional socioeconomic climate (Serebryakova et al., 2020).

Since the 90's, literature highlighted the significance of the interactions among the people and the institutions involved in innovation; indeed, interactions among innovation actors produce systemic innovation (Cooke, 2001) and have positive influences for the economic development of related community (Woolcock, 1998); accordingly, the concept of innovation system emerged in order to understand the innovation dynamics in a given system, acting as an analytical tool in order to identify the determinants of its systemic innovation, and in order to formulate and assess possible strategies focused on a territorial scale (Tödtling et al., 2022), with the final aim to define the relevant framework to be considered for the policy designing.

David and Metcalfe (2008) suggested that the term "innovation systems" is misleading because it emphasizes static and durable institutional structures. Indeed, under the development of "innovation systems" there are emergent properties of an ecology of innovation, resulting from the formation of mutually reinforcing inter-organizational relationships between individual and organizational entities specialized in functional capabilities. Within the above conceptual framework, <<nonlinear models of innovation have been introduced to take interactive and recursive terms into account.>> (Hajek & Henriques, 2017).

Literature highlighted geographical proximity, among innovation actors, as a fundamental element for the development of systemic innovation; indeed, as argued by Boschma R.A. (2005), geographical proximity of innovation agents is a necessary prerequisite for the innovation development because it facilitates interpersonal contacts, information exchange and trust; accordingly, Kirat and Lung (1999) argued that geographical proximity promotes collective learning processes; moreover, geographical proximity helps exchange of the key elements of innovative activities such as tacit knowledge and grey knowledge (Asheim & Gertler, 2005; Iandoli & Zollo, 2007), and facilitates interpersonal contacts, interactive and collective learning (Kirat & Lung, 1999); furthermore, the geographical proximity is able to foster the reinforcement of the other proximity dimensions as well as the cognitive proximity (the sharing of a common knowledge and competence base), the organizational proximity (the sharing of a common capacity to coordinate and exchange the knowledge), the social proximity (the sharing of social ties of friendship and trust), and the institutional proximity (the sharing of the same institutional rules of the game like a set of cultural habits and values) (Boschma R.A., (2005).

Literature also highlighted the critical role played by the local institutional conditions for their impact onthe systemic development of the innovation capability; indeed - according to Boschma R.A. (2005) - institutions are able to balance every form of proximity (cognitive, organizational, social, institutional, geographical); more in general, institutions are able to shape the behavior of the innovation actors - and their relationships - with laws, regulations, values, practices, routines (Trippl, 2006; Andersson & Carlssons 2006).

From the point of view of the geographical dimension of the innovation, Cooke and Memedovic (2003) argued that regions, especially those with established clusters and effective administrative structures to foster innovative businesses, serve as more significant economic communities because they can <<define genuine flows of economic activities and can take advantage of true linkages and synergies among economic actors.>>. Moreover, literature highlighted that the accumulation of technological processes occurs mainly on the regional level and highlighted that the technological and knowledge spillovers tend to be geographically concentrated (Brenner & Grief, 2006. OECD, 2011).

The smaller is a territory the bigger is the level of geographical proximity of innovation actors, and the bigger is a territory the higher is the number and variety of innovation actors; therefore, it could be argued that geographical proximity and critical mass/variety of innovation actors go in the opposite way with respect to the territorial dimension; so, according to the abovementioned literature on regional dimension of innovation, it may be inferred that regional dimension represents the ideal intermediate dimension, between too small and too big territories - for balancing the needs of the geographical proximity and the needs of the critical mass/variety of the innovation actors. In other words, it can be stated that regional level works better - with respect to the innovation development - than national level or sub-regional level.

As examples of empirical evidence on the relevance of the regional dimension for systemic innovation, it is well known that there has been a polarization of innovation in certain regions around the world (i.e.: Silicon Valley, Baden Wurttemberg); this phenomena has been linked to spatial agglomeration, face-to-face interactions and exchange of tacit knowledge (Pinheiro et al., 2022), and is related to institutional condition and to the presence of intangible resources - such as culture, competence, and knowledge - at regional level. Supporting this view, Tödtling and Trippl argue that relationship between innovation actors is largely influenced by the regional institutional asset; on this point they suggest that regional dimension of knowledge creation process represents an effective tool for understanding regional disparities in innovative capacity (Tödtling et al., 2022; Tödtling & Trippl, 2012).

Furthermore, on the above point, it is important to emphasize how challenges related with innovation and economic and social growth are mainly conditioned by intellectual potential of a territory (Serebryakova et al., 2020).

Since 2002 the European Commission carried out publications with data and statistical evaluations on the innovative capability of the EU regions; moreover, the above publications fostered the development of periodical innovation scoreboards related to European regions (Union Regional Innovation Scoreboard) (Hollanders & Es-Sadki, 2021a; Hollanders & Es-Sadki, 2023a). Last ones publications are characterized by the adoption of specific performance indicators and by the implementation of a composite indicator (Regional Innovation Index – RII) which is aimed at measuring the multidimensional concept of the regional innovation capability.

In the European area this method is applied to NUTS 2 regions, where NUTS referring to territorial classifications of <<Nomenclature des Unités Territoriales Statistique>> (Nomenclature of statistical territorial units); NUTS differentiation is based on different territorial levels: States (NUTS 0), macro areas (NUTS 1), Regions (NUTS 2) and provinces (NUTS 3).

To date, regional innovation capability is periodically measured, in the European Union (EU), through the average of several innovation indicators used in the regional innovation scoreboard (Hollanders & Es-Sadki, 2021a; Hollanders & Es-Sadki, 2023a).

The European Union Regional Innovation Scoreboard is broadly adopted by European policymakers as a classification tool and a ranking one (Hollander & Es-Sadki, 2021b; Hollanders & Es-Sadki, 2023b), because of the presence of the regional innovation indicators tables, the classification of regions with respect to the relative innovation performances and the regional innovation performance ranking (Hollanders & Es-Sadki, 2021a; Hollanders & Es-Sadki, 2023a). The aforementioned document reports the value - over time - of the innovation indicators of each region; these over time values can, of course, be useful for analyzing the evolution of each specific regional indicator of innovation, from the past to the present.

The European Regional Scoreboard is broadly used for the definition of regional innovation policy by European policymakers; but we highlight that European regional innovation scoreboards are characterized by a certain grade of methodological weakness because of the absence of a conceptual framework beyond, and by the presence of some disadvantages linked to the related taxonomic method behind (Szopik-Depczyńska et al., 2020); we also highlight that scoreboards are very useful tools because they let to measure the level of local resources and competencies; thus, we argue the importance of the European Union Regional Innovation Scoreboard as a regional innovation measuring tool for the analysis - over time - of each specific indicator, but we also argue that the European regional policymakers need integrative methodological tools aimed at performing what-if analysis in order to define the most effective regional innovation policies to implement.

To best of our knowledge, literature lacks contributions aimed at integrating effective policy scenario tools to the European Regional Innovation Scoreboard. Thus, to fill this gap, section 2 presents the relevant background aspects related to the topic; the research questions are presented in section 3; the methodology is reported in section 4; section 5 presents the discussion of the results; finally, conclusions are presented in section 6.

# 2. Background

**2.1 European Regional Innovation Scoreboard**

Innovation system can be viewed as a complex system of firms, knowledge actors and institutions that impact economic and innovative performance within a region (Asheim & Gertler, 2005). According to Cooke, the innovation system is based on the establishment of a systemic and cooperative context that strongly depends on relationship of mutual trust between actors that take part of it (Cooke et al., 1997). According to Tödtling and Trippl, innovation system has to be seen as an evolutionary, non-linear, intense and interactive process between different actors (universities, innovation centers, education institutions, financial institutions, standard setting organisms, trade association and government agencies) (Tödtling & Trippl, 2005).

Regions are the most correct level for the definition of effective policies which are able to improve the interactions between innovation actors and, as consequence, which are able to get better the innovation capability of the related innovation system. By now, due to the territorial dimension of innovation processes, to the importance of local institutional conditions and - still - to the systemic character of the cooperation and the mutual learning among innovation actors (Tödtling et al., 2022), the Regional Innovation System (RIS) is considered the most adequate conceptual framework aimed at defining the most effective policies which can improve the innovative capability of a territory.

RISs are complex systems resulting from the integration of a territorially embedded institutional infrastructure and a production system (Doloreux, 2002). The RIS framework defines the innovation as a cumulative and not-linear systemic process (Fischer, 2001) resulting from the formal and informal, voluntary and involuntary interactions between different actors operating in the innovation system.

The main idea - in the RIS approach - is that interactions, among different local actors that have good reasons to interact (such as small and large firms, manufacturing and service companies, industries and universities, private and public agencies), should foster local learning processes.

In literature, several assessment techniques of regional innovation capability have emerged. The main group of methods, aimed at assessing the level of innovation, are statistical and mathematical methods that are based on various indicators and measures. In particular for the European regional innovation capability, the European Union periodically has developed the European Regional Scoreboard as a statistical tool for classification and ranking purposes of the European regions (Hollanders & Es-Sadki, 2021a; Hollanders & Es-Sadki, 2023a).

The European Union Regional innovation scoreboard (EURIS) is usually used as a reference guide by the innovation policy makers across the EU. The EURIS measures - on the periodic basis - the innovation performances of the European regions by evaluating a set of innovation indicators. Average innovation performance is then measured by using a composite indicator calculated as the un-weighted average of the normalized scores of the above indicators (Hollanders & Es-Sadki, 2021b; Hollanders & Es-Sadki, 2023b); the final scoreboard classification - of each region - is grounded on a composite indicator that is not related to any specific conceptual model of innovation; thus, as consequence – the final scoreboard classification is characterized by a certain degree of methodological weakness because of the absence of a conceptual framework beyond.

The EURIS classifies regions - and elaborates regional ranking - by means of an un-weighted average of innovation indicators (Hollanders & Es-Sadki, 2021b; Hollanders & Es-Sadki, 2023b), but we highlight that the relative methodology doesn't take in account the possible correlation among innovation indicators; as consequence, we highlight that region's classification and region's ranking could be biased because of the eventual presence of correlated indicators that can be linked to the same information and that can push the un-weighted average towards biased values.

Moreover, the above adoption of the un-weighted average of the innovation indicators does not take into account the processes - not linear and complex – underlying the development of the regional innovative capacity, and does not take into account the consequent need to operate according to non-linear models, as suggested by the literature on the subject (Hajek & Henriques, 2017). But we also highlight that clustering methodologies - that operate according to the similarity of the entire set of indicators, and that cluster regions non-linearly - lend themselves well suited to the modeling of regional innovative capacity.

Thus, there is the need to develop a synergistic clustering tool - of the European regional Innovation Scoreboard - that is aimed at taking in account the non linearity and complexity of the regional innovation development, according to the idea << to look for regions similar to each other due to groups of indicators>> (Szopik-Depczyńska et al., 2020); such a clustering methodology should be able to overcome the information redundancy issue that is linked to the possible correlations among innovation indicators; furthermore, such a methodology should be able to work on uncorrelated innovation drivers by reducing the dimensionality, of original innovation variables (i.e.: by developing new synthesis variables that are orthogonal to each other, and, therefore, that are not correlated by construction).

Several attempts to apply different methodologies for the clustering and classification of the regional innovative capacity have emerged in the literature: multi-criteria taxonomy method (Szopik-Depczyńska et al., 2020), linear ordering method based on Hellwig's synthetic measure (Bielinska-Dusza & Hamerska, 2021), Multicriteria group decision analysis (Paredes-Frigolett et al., 2021). The above attempts, however, frequently do not operate the appropriate reduction of the dimensionality mentioned above; the statistical literature, on the other hand, highlights the existence of clustering methodologies which are also based on the reduction of dimensionality.

According to the need of taking in account the non linearity and complexity of the regional innovation development, and according to the idea « to look for regions similar to each other due to groups of indicators» (Szopik-Depczyńska et al., 2020), we claim the need to define a new labeling methodology – for innovation capability labeling of European regions - which could be based on an unsupervised learning method. But, generally speaking, a primary challenge in unsupervised learning arises from the lack of predefined labels; as consequence, understanding the meaning and significance of clusters extracted from unsupervised learning can be a complex task; in more details on this point, literature highlights that labeling methods are essentially divided into two main research fields: labeling data methods based on using existing labeled data, and labeling data methods based on not using labeled data; each one of the above two general methodologies could be automated within the machine learning framework, in order to gain speed and objectiveness: with respect to automatic methodologies for labeling purposes, based on using existing labeled data, Sager et al. (2021) have developed a survey in which they highlighted the literature mainstream related to the leverage of existing labeled data to generate annotation automatically. According to the need to develop automatic labels for numerical data without using existing labeled data – such as in the addressed problem of regional innovation labeling in which there aren't ground labeled truth data - literature highlighted the need to use clustering-based labeling methodologies to group similar data points, and then to assign labels, based on cluster characteristics, by extracting important features (with Principal Component Analysis or with some other methodology based on feature importance analysis); as example of clustering-based labeling methodologies based on feature importance analysis, Shaheen et al. (2013) proposed a method to label unsupervised classes - based on clustering - in which they assign each analyzed statistical unit (a country) to a labeled cluster and in which labels are assigned with a feature importance analysis based on the correlation value found between original indicators (i.e.: original features) and two specific variables (Production rate and consumption rate). With respect to the addressed problem of regional innovation grouping, we further highlight that clustering methodologies - that are based on the similarity of the entire set of indicators, and that are able to cluster regions non linearly (thus, that are in accordance with the need of taking into account the processes, not linear and complex, underlying the development of the regional innovative capacity) - let themselves well suited to the modeling of regional innovative capacity. With respect to the above theoretical framework, we further claim the need to develop a synergistic clustering tool - of the European regional Innovation Scoreboard - that is aimed at overcoming the information redundancy issue that is linked to the possible correlations among innovation indicators (i.e.: according to the need of overcoming the EURIS methodological issue of the correlated indicators usage); such a methodology should be able to work on uncorrelated innovation drivers by means of some technique that is able to develop new synthesis variables that are orthogonal to each other (and, therefore, that are not correlated by construction); last - but not least – such a methodology should be able to label the clusters without using data labels, in accordance with the absence of a ground truth on the original dataset of the EURIS scoreboard, but by using a feature importance analysis technique - to assign labels to every cluster - in accordance to literature (Shaheen et al., 2013). Thus, with the aim to group and to label regions with respect to regional innovation capability, we claim that there is the need to use methodologies of joint dimension reduction and clustering of data, which are aimed at preserving as much of the data's variability as feasible, while reducing it to a minimal number of uncorrelated dimensions, and which are also able to give us the opportunity to define new labels for the original EURIS data;

# 3. Research questions

We highlight that the European Union Regional Innovation Scoreboard is a very useful tool for analyzing the overtime evolution of each specific regional innovation indicator; but we also highlight that the un-weighted average of the innovation indicators - that is used for the calculation of the EURIS composite indicator and which is the criterion on which regions' classification and regions' ranking is based on - doesn't take in account the possible correlation among innovation indicators; as consequence, we highlight that EURIS region's classification and region's ranking could be biased because of the eventual presence of correlated indicators that can be linked to the same information and that can push the un-weighted average towards biased values.

Moreover, from the modeling point of view an un-weighted average of the innovation indicators does not take into account the processes - not linear and complex – underlying the development of the regional innovative capacity, and does not take into account the consequent need to operate according to non-linear models

As the European Regional Innovation Scoreboard is currently and broadly considered the reference document for the definition of regional innovation policies by European policymakers, according to our above view we highlight the need to integrate the EURIS scoreboard with a synergistic tool aimed at helping policymakers for the development of optimal innovation policies. So, according to our idea, the development of new operational models, and their synergistic use with the European Regional Innovation Scoreboard, should be addressed to answer all the following further questions:

1) Which is the belonging cluster of each region?
2) What are the most effective policies to be implemented for each region?

# 4. Methodology

**4.1 Determinants of the Regional Innovation Performance of the EURIS 2023**

In this study, we use the indicators of the European Union Regional Innovation Scoreboard 2023 (Hollanders & Es-Sadki, 2023a) presented in Table 1:

| INNOVATION DETERMINANT 1: FRAMEWORK CONDITIONS | |
|---|---|
| **Innovation Dimensions** | **Indicators** |
| Human resources | Percentage of population aged 25-34 having completed tertiary education |
| | Lifelong learning, the share of population aged 25-64 enrolled in education or training aimed at improving knowledge, skills and competences |
| Attractive research systems | International scientific co-publications per million population |

|  | Scientific publications among the top-10% most cited publications worldwide as percentage of total scientific publications of the country |
|---|---|
| Digitalisation | Individuals with above basic overall digital skills |

**INNOVATION DETERMINANT 2: INVESTMENTS**

| Innovation Dimensions | Indicators |
|---|---|
| Finance and support | R&D expenditure in the public sector as percentage of GDP |
| Firm investments | R&D expenditure in the business sector as percentage of GDP |
|  | Non-R&D innovation expenditures as percentage of total turnover |
|  | Innovation expenditures per person employed in innovation-active enterprises |
| Use of information technologies | Employed ICT specialists |

**INNOVATION DETERMINANT 3: INNOVATION ACTIVITIES**

| Innovation Dimensions | Indicators |
|---|---|
| Innovators | SMEs introducing product innovations as percentage of SMEs |
|  | SMEs introducing business process innovations as percentage of SMEs |
| Linkages | Innovative SMEs collaborating with others as percentage of SMEs |
|  | Public-private co-publications per million population |
| Intellectual assets | PCT patent applications per billion GDP (in Purchasing Power standards) |
|  | Trademark applications per billion GDP (in Purchasing Power standards) |
|  | Individual design applications per billion GDP (in Purchasing Power standards) |

**INNOVATION DETERMINANT 4: IMPACTS**

| Innovation Dimensions | Indicators |
|---|---|
| Employment impacts | Employment in medium-high and high-tech manufacturing and knowledge-intensive services |
|  | Employment in innovative enterprises |
| Sales impacts | Sales of new-to-market and new-to-enterprise product innovations as percentage of total turnover |
| Environmental sustainability | Air emissions in fine particulates (PM2.5) in Industry |

Table 1: indicators of the European Union Regional Innovation Scoreboard 2023

We adopt the NUTS 2 regions included in the European Union Regional Innovation Scoreboard 2023 (Hollanders & Es-Sadki, 2023a), where the NUTS classification (Nomenclature of territorial units for statistics) is a structured system used to partition the European Union's economic territory into different levels; it consists of three tiers: NUTS 1, which identifies major socio-economic regions; NUTS 2, designating basic regions for implementing regional policies; and NUTS 3, which defines smaller regions for specific analysis and assessment.

## 4.2 Methodologies

**Clustering methodologies**

Clustering methodologies are well suited to the modeling of regional innovative capacity because these ones are able to develop regional clusters nonlinearly, according to the similarity of the entire set of indicators, in order to << to look for regions similar to each other due to groups of indicators>> (Szopik-Depczyńska et al., 2020).
An effective clustering methodology - to be choose - should be able to overcome the information redundancy issue that is linked to the possible correlations among original innovation indicators; accordingly, such a methodology should be able to reduce the dimensionality, of original innovation variables, by developing new synthesis variables that are orthogonal to each other, and, therefore, that are not correlated by construction.

Dimension reduction and cluster analysis are frequently employed - as tandem approach (Markos et al., 2019) - by executing them in sequence: first of all, the original data is reduced in dimension, and subsequently, cluster analysis is developed on the data of the reduced sub-space. By means of this approach, results are unaffected by multicollinearity (Hájková & Hájek, 2010).

While the tandem approach is easy to understand and apply, this methodology cannot be able to develop the best cluster assignments because of different objectives that are optimized by the two component methods; indeed, dimension reduction is aimed at preserving the maximum data variance within a reduced dimensionality, whereas cluster analysis is aimed at identifying similarities among statistical units and is aimed at grouping them accordingly. This discrepancy in goals is acknowledged as a known concern (De Soete & Carroll, 1994; Van Buuren & Heiser, 1989; Vichi & Kiers 2001), and several proposed solutions - that integrate dimension reduction and clustering - have been suggested in literature; for continuous data, literature suggested methods such as reduced K-means (De Soete & Carroll, 1994), factorial K-means (Vichi & Kiers 2001), and a related hybrid approach. For categorical data, literature developed cluster correspondence analysis (van de Velden et al., 2017), multiple correspondence analysis and K-means in a unified framework (Hwang et al., 2006), as well as iterative factorial clustering of binary variables (Iodice D'Enza & Palumbo 2013; Markos et al., 2019).

Factorial K-means (FKM) and reduced Kmeans (RKM) are clustering methodologies, to be used on continuous data, that are well suited for the aim to find clusters that are based on a complex set of non redundant information (i.e. on a set of non correlated variables that are able to contribute, in a non linear manner, to a latent effect); more on this point, FKM and RKM methodologies are well suited to our task because:

1) FKM and RKM methodologies group the regions according to similarity criteria, by taking into account, therefore, the information contained in the whole set of indicators; more on this point with respect to the topic of this paper, we highlight that the overall information extracted from the whole set of indicators is consistent – from the theoretical point of view - with the concept of non-linearity and complexity that characterizes the development of regional innovative capacity. It is also important to highlight that FKM and RKM methodologies group the regions according to the distance of each region with respect to the centroid of the belonging cluster, in accordance with our research questions.

2) FKM and RKM methodologies work on uncorrelated latent variables, as they operates a reduction in the dimensionality of the original statistical variables (i.e. the original innovation indicators) in latent variables that are orthogonal to each other and, therefore, that are uncorrelated by construction.

In essence, FKM and RKM perform a cluster analysis jointly with Factorial Analysis, and work not on the original variables but on latent ones which are uncorrelated by construction.

As stated in De Soete and Carroll (1994), RKM minimizes the distance between the statistical units from the original space and the 'quasi' centroids of the reduced subspace; on the other hand, FKM is aimed at minimizing the total squared distances between the centroids in the projected space and the data points projected onto the same subspace where these centroids are situated. (Timmerman et al., 2010)

Markos et al. (2019) stated that the RKM and FKM are closely related. Furthermore, literature revealed how the performance of FKM and RKM methods is influenced by the existence of residuals within the clustering subspace and/or within its orthogonal complement: in the case of FKM, as the proportion of residuals within the subspace becomes larger with respect to the residuals in the complement, the result quality get worst. On the contrary for reduced K-means, when the proportion of residuals in the complement becomes larger than those in the subspace, the result quality gets better. From another point of view, Reduced K-Means works well, and Factorial K-Means does not work when the subspace containing the cluster structure has more variability than the orthogonal subspace. In contrast, Factorial K-Means performs well, and Reduced K-Means does not perform, when the data exhibit much variability in directions that are orthogonal to the subspace containing the cluster structure (Terada, 2014). Thus, FKM and RKM exhibit a complementary relationship, each addressing the issues of the other in distinct scenarios (Timmerman et al., 2010)

In RKM, the simultaneous dimension reduction and cluster analysis problem is addressed by a cluster allocation that maximizes the "between" variance of the clusters in the reduced space (Markos et al., 2019); on the other side, in FKM the jointly dimension reduction and cluster analysis problem is addressed by a cluster allocation that minimizes the "within" variance of the clusters in the reduced space (Markos et al., 2019). As we deal with continuous data, and as we are interested to develop clusters that aggregate regions according to the maximum similarity criterion (i.e. with the minimum "within" variance of the clusters in the reduced space), we choose Factorial K-means (FKM) as clustering methodology of this paper.

By means of this choice, we could also develop an intra-cluster regional ranking among all regions belonging to each specific cluster (i.e.: among all the regions that present the maximum similarity with respect to the overall non linear effect - of the whole set of innovation indicator values - on the regional innovation capability); the above intra-cluster ranking can be developed by estimating – within each cluster – the distances between the innovation capability of each region (belonging to the considered cluster) and the target innovation capability of the considered cluster, where the above target has to be defined according to the semantic meaning of the latent variables of the reduced space; on last point, it is important to note that latent variables represents levels of latent innovation macro-drivers that need to be named according to a specific factorial analysis.

Furthermore, we highlight that a linear combination of latent variables can be used as innovation index for regional innovation ranking purposes, because latent variables - that are uncorrelated by construction (and, as consequence, that are linked to different information contents, by construction) - can be conceptually linearly added without potential correlation-based bias, differently from a possible bias that could be linked to an un-weighted average of (possible correlated) innovation indicators.

By this methodology we could also evaluate the "first in class" region, of each specific cluster ranking, as first target innovation capability level to be addressed from each region belonging to the same cluster; indeed, "first in class" region - of each specific cluster - represents a concrete example of regional innovation capability that could be reached by each one of the regions belonging to the same cluster, as the "first in class" region" is a similar region from the point of view of the whole set of innovation indicators, i.e. from the point of view of the whole innovation regional structure.

After defining the names of latent innovation macro-drivers (by means of the analysis of original variables coordinates on latent variables), for clusters' labeling purposes we analyze centroid's coordinates of every cluster, with respect to latent variables, in order to assign a label to each cluster according to EURIS original names, as detailed in section 4.4; we use EURIS original names - for our cluster labeling process - in order to go toward a synergistic use of our tool with respect to the European regional Innovation Scoreboard.

**Neural network methodologies**

Empirical findings have highlighted that relationship between innovation performance and innovation drivers is complex and non linear (Hajek & Henriques, 2017).

In traditional regression models the predictions accuracy can be compromised in dynamic, noisy, and volatile settings; in contrast, Artificial Neural Networks (ANNs) possess the ability to naturally capture complex and non-linear connections between inputs and outputs. More specifically with respect to the topic of this paper, ANNs have proven their effectiveness in forecasting technology implementation results for the research and development performance of European nations (Hajek & Henriques, 2017).

As stated in previous sections, an useful methodological tool - for regional innovation policymakers - should give the opportunity to simulate the most effective policies to be implemented, for each region, in order to reach a target result; thus, as the ANN's capability to capture complex non-linear connections between regional innovation drivers (inputs) and regional innovation capability (outputs), and according with several literature suggestions (Hajek & Henriques, 2017; de la Paz-Marín et al., 2012; Pei et al., 2022), we use the neural network methodology for a what-if analysis aimed at predicting the effect of regional innovation policies, with the final aim to understand what is the policy path to be addressed - by each region – in order to push itself into a stronger innovation cluster.

ANN is able to predict the innovation cluster label assignment for every specific set of innovation indicators values; thus, by means of ANN we are able to understand if specific regional policies (i.e.: specific set of regional innovation indicators values) can give positive contribution to the improvement of regional innovation capability, or not; thus, the above what-if methodology can provide regional policymakers with the opportunity to predict the most effective policy path for promoting regional innovation.

### 4.3 Dataset description

We use the data of European Union Regional Innovation Scoreboard 2023 (Hollanders & Es-Sadki, 2023b), in accordance with the following classification scheme:

- Innovation Leaders: regions with a relative performance more than 125% of the EU average in 2023
- Strong Innovators: regions with a relative performance between 100% and 125% of the EU average in 2023
- Moderate Innovators: regions with a relative performance between 70% and 100% of the EU average in 2023
- Emerging Innovators: regions with a relative performance below 70% of the EU average in 2023

As the EURIS 2023 dataset includes regional data also belonging to previous years, we elaborate our analysis by considering the overall picture for all the EURIS regions – from 2016 to 2023; as result, we have a dataset constituted by 1912 observations; as the whole period includes lots of missing data for 7 indicators, we develop our work by considering only 14 indicators – selected from the ones of table 1 – according to the idea to deal with indicators without missing data. Figure 1 presents the selected indicators.

| 1.1.2 Population with tertiary education | 1.1.3 Population involved in lifelong learning | 1.2.1 International scientific co-publications | 1.2.2 Scientific publications among the top 10% most cited | 1.3.2 Individuals with above basic overall digital skills | 2.1.1 R&D expenditure in the public sector | 2.2.1 R&D expenditure in the business sector |
| --- | --- | --- | --- | --- | --- | --- |
| 2.3.2 Employed ICT specialists | 3.2.2 Public-private co-publications | 3.3.1 PCT patent applications | 3.3.2 Trademark applications | 3.3.3 Design applications | 4.1.1 Employment in knowledge-intensive activities | 4.3.2 Air emissions by fine particulates |

Figure 1: selected indicators for our analysis

Furthermore, we assign to each label of EURIS classification a numeric code, according to table 2

| Label of EURIS | Assigned number |
| --- | --- |
| Innovation leaders | 1 |
| Strong Innovators | 2 |
| Moderate innovators | 3 |
| Emerging innovators | 4 |

Table 2: numeric code assigned to each EURIS label

## 4.4 Results

In section 2.1 we highlighted that the EURIS adoption of the un-weighted average (of the innovation indicators), as criterion for regional classification and regional ranking, should not be used from the theoretical point of view because the linear model behind the un-weighted average is not conceptually coherent with the process - non linear and complex – underlying the development of the regional innovative capacity; furthermore, in section 2.1 we also highlighted that the EURIS classifies regions - and elaborates regional ranking - by means of the un-weighted average of possible correlated innovation indicators; if innovation indicators are correlated then EURIS region's classification and EURIS region's ranking could be biased because correlated indicators are linked to the same information and could push the un-weighted average towards biased values. Thus, in order to develop a synergistic tool of the European Regional innovation

Scoreboard that is able to overcome the potential conceptual issues of above, we start our analysis by checking possible correlation among original innovation indicators of the EURIS 2023; from this point of view, our results highlights the presence of a significative level of correlation between some indicators, according to figure 2 and to annex 1 results.

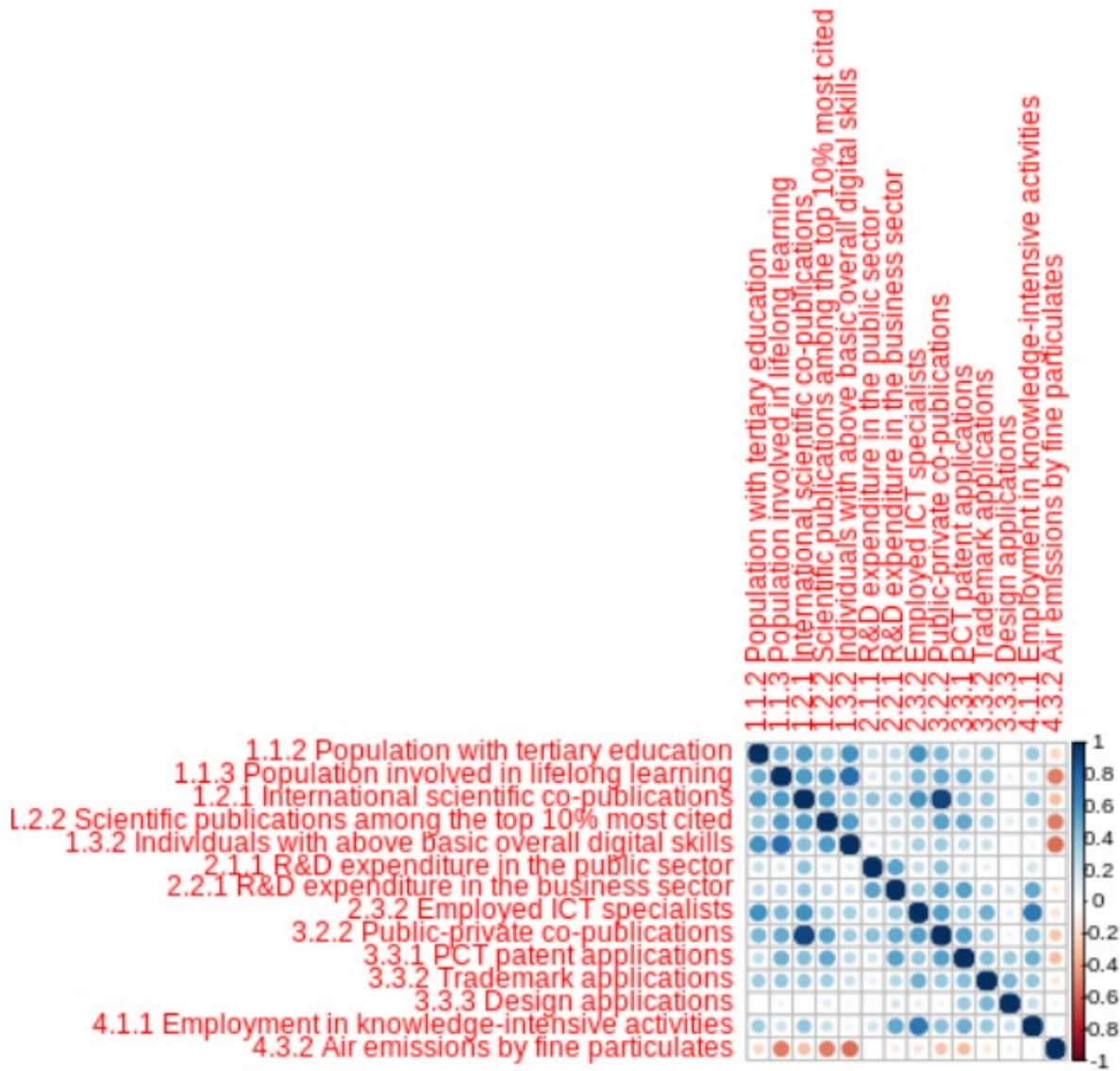

Figure 2: correlation matrix

Thus, in order to develop an EURIS synergistic tool for regional innovation policymakers, as highlighted in section 2.1 and 4.2, we propose a methodology of joint dimension reduction and clustering of data with the aim to get results in a "non linear" manner, and with the scope to overcome the information redundancy issue linked to the correlation among original innovation indicators.

Our elaboration starts by developing a Principal Component Analysis (PCA) with the aim to know the number of latent variables to be used as input parameters of the Factorial K-Means. The optimal number of latent variables is selected according to the "elbow" rule with respect to the principal components with eigenvalue >1 (in accordance with the other general rule that a principal component, with eigenvalue <=1, hasn't to be considered because it is less informative than the original standardized variables); this rule identifies the point of the "elbow" on a plot between eigenvalues (on the y-axis) and the number of principal components (on the x-axis); so, this rule suggests to select all components before the rate of decrease in eigenvalues slows down significantly (i.e.: before the line flattens out); indeed, the principal components before this point contain the most meaningful information, while those afterward contribute relatively little; in other words, principal components following the drop in eigenvalue have to be excluded because they add relatively little to the information already extracted by the principal components before the drop. Accordingly, as the eigenvalue divided by the sum of the eigenvalues is the proportion of variance explained by the relative principal component, with the aim to select the optimal number of principal components, we can also apply the "elbow" rule on the proportion of variance explained by every principal component with eigenvalue > 1. Results of PCA - with respect to the proportion of variance explained by every principal component - are presented in Table 3 and in figure 3a and 3b:

|  | Comp.1 | Comp.2 | Comp.3 | Comp.4 | Comp.5 |
|---|---|---|---|---|---|
| Standard deviation | 2.4000200 | 1.4000605 | 1.1864626 | 1.05421891 | 0.86343132 |
| Proportion of Variance | 0.4114354 | 0.1400121 | 0.1005495 | 0.07938411 | 0.05325097 |
| Cumulative Proportion | 0.4114354 | 0.5514475 | 0.6519971 | 0.73138118 | 0.78463215 |
|  | Comp.6 | Comp.7 | Comp.8 | Comp.9 | Comp.10 |
| Standard deviation | 0.85429454 | 0.69268053 | 0.65337672 | 0.61366705 | 0.59361866 |
| Proportion of Variance | 0.05212994 | 0.03427188 | 0.03049294 | 0.02689909 | 0.02517022 |
| Cumulative Proportion | 0.83676209 | 0.87103397 | 0.90152691 | 0.92842600 | 0.95359622 |
|  | Comp.11 | Comp.12 | Comp.13 | Comp.14 |  |
| Standard deviation | 0.53747925 | 0.41062458 | 0.367144738 | 0.239501814 |  |
| Proportion of Variance | 0.02063457 | 0.01204375 | 0.009628233 | 0.004097223 |  |
| Cumulative Proportion | 0.97423079 | 0.98627454 | 0.995902777 | 1.000000000 |  |

Table 3: proportion of variance explained by every principal component

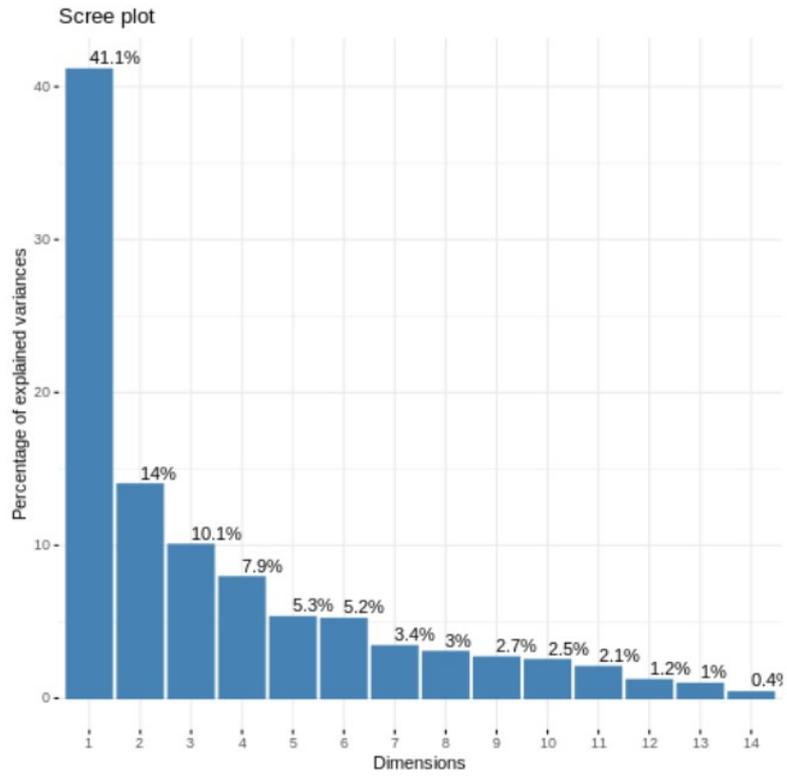

Figure 3a – percentage of explained variances Vs principal components

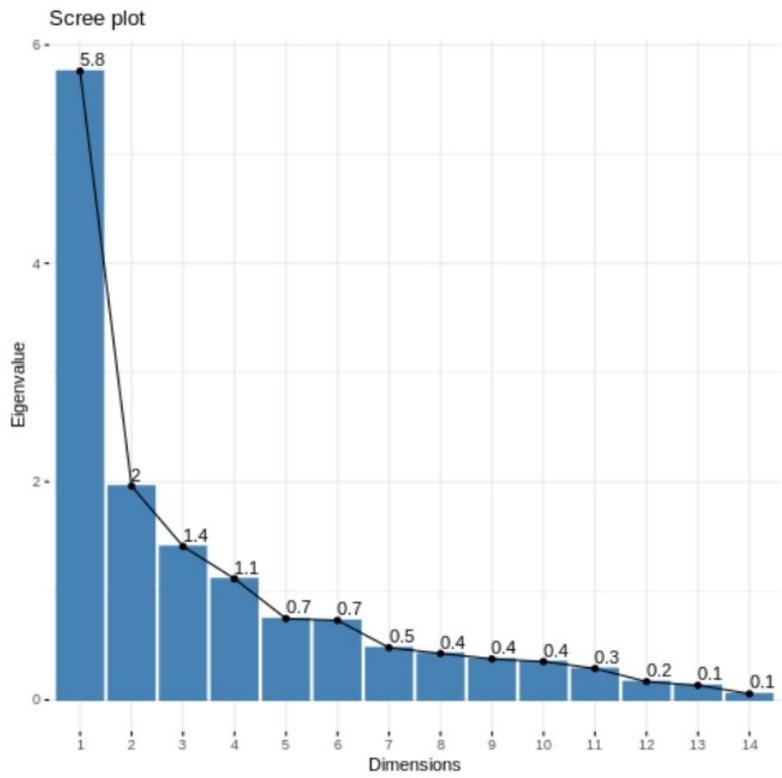

Figure 3b – eigenvalues Vs principal components

By considering the results of table 3, figure 3a and figure 3b, we choose only first and second principal components as latent variables, because the jump in explained variance between second and third component is not relevant when compared to the jump in explained variance between first and second component. First 2 principal components explain 55,14% of data variance on the analyzed data. In annex 1 we also present detailed results of PCA.

When data have similar variation ranges and same units of measurements, covariance matrix is the optimal matrix choice to perform a PCA because covariance matrix let to preserve variance in the data, or – in other words - let to preserve the amount of information of the dataset; indeed, one way to think about the amount of information into a dataset is to see how spread out the data points are, as no spread (or variability) in the data means little information (with no spread the knowledge of one data value let us to know the other ones); but in current case the original variables have different ranges of variation and different units of measurement; thus, we carried out the principal component analysis by means of zero-mean standardization of the original data, i.e. by centring them on the relative average value and by scaling them to unit variance; in accordance with this view, we have performed the PCA on the correlation matrix, as it is equivalent to deal with zero-mean standardized variables.

The use of correlation matrix, notwithstanding the loss of information due to the flattening of variance, can give useful insights on the relationship among variables (original variables and/or latent variables); in fact, scores of original variables on principal components let to interpret the meaning of latent variables: the greater are the scores of original variables in absolute value, the greater is the impact of the relative original variable on the principal components (with a direct relationship or an inverse one, depending on the sign of the score).

According to the above overall results, for the jointly dimension reduction and the clustering process we develop a Factorial K-means (detailed results are also presented in Annex 2) by respectively considering 2 factors and 4 clusters; we highlight that our 4 clusters choice is aimed at identifying a number of regional groups that is coherent with a synergistic use of our proposed tool with the European regional Innovation Scoreboard.

Table 4-a presents scores of original variables with respect to first and second principal component, as results from the "cluspca" function (https://www.rdocumentation.org/packages/clustrd/versions/1.4.0/topics/cluspca), which is the one that implements Factorial K-means in R coding language.

First principal component highlights greater scores for <<2.1.1 R&D expenditure in the public sector>> with plus sign, and <<2.2.1 R&D expenditure in the business sector>> with plus sign; second principal component highlights greater scores – with minus sign - for all of the following innovation indicators: <<1.1.3 Population involved in lifelong learning>>, <<1.2.2 Scientific publications among the top 10% most cited>>, <<1.3.2 Individuals with above basic overall digital skills>>, <<3.3.1 PCT patent applications>>, <<3.3.2 Trademark applications>>.

```
Variable scores:
                                                          Dim.1   Dim.2
1.1.2 Population with tertiary education                 -0.1814 -0.0848
1.1.3 Population involved in lifelong learning           -0.1226 -0.3924
1.2.1 International scientific co-publications            0.1413 -0.2008
1.2.2 Scientific publications among the top 10% most cited  0.0260 -0.4246
1.3.2 Individuals with above basic overall digital skills -0.1981 -0.3416
2.1.1 R&D expenditure in the public sector                0.6283  0.0470
2.2.1 R&D expenditure in the business sector              0.5583 -0.0758
2.3.2 Employed ICT specialists                           -0.0916 -0.0013
3.2.2 Public-private co-publications                      0.2782 -0.2327
3.3.1 PCT patent applications                             0.2617 -0.3602
3.3.2 Trademark applications                             -0.1005 -0.3239
3.3.3 Design applications                                -0.0125 -0.2420
4.1.1 Employment in knowledge-intensive activities        0.1127  0.0894
4.3.2 Air emissions by fine particulates                  0.0920  0.3718
```
Table 4-a: scores of original variables with respect to first and second principal component

According to table 4-a, regions that spend a lot on R&D (in the public and/or in the business sector) have positive and high values for the first principal component; thus, essentially, the first main component can be interpreted as the level of R&D spending. The second principal component is inversely linked to the technical-scientific capabilities of individuals belonging to a region: regions whose individuals have low technical-scientific capacity will have a high value of the second principal component; therefore, the second principal component can be interpreted as the technical-scientific "lag" of individuals belonging to a region.

An extract of overall results (EURIS results = <<Original_label (1_innovation_leader; 2=strong; 3=moderate; 4=emerging)>> and Factorial K-means results = <<output_ cluster from ClusPca package>> ) is presented in table 4-b.

| Region | 1.1.2 Population with tertiary education | 1.1.3 Population involved in lifelong learning | 1.2.1 International scientific co-publications | 1.2.2 Scientific publications among the top 10% most cited | 1.3.2 Individuals with above basic overall digital skills | 2.1.1 R&D expenditure in the public secto | 2.2.1 R&D expenditure in the business sector | 2.3.2 Employed ICT specialists | 3.2.2 Public-private co-publications | 3.3.1 PCT patent applications | 3.3.2 Trademark applications | 3.3.3 Design applications | 4.1.1 Employment in knowledge-intensive activities | 4.3.2 Air emissions by fine particulates | Original_label(1_innovation_leader;2=strong;3=moderate;4=emerging;) | output_cluster from ClusPca package |
|---|---|---|---|---|---|---|---|---|---|---|---|---|---|---|---|---|
| 1 | 47,1 | 16,2 | 3078 | 11,14 | 33,51 | 1,23 | 1,57 | 6,5 | 620,7 | 3,275 | 12,91 | 4,041 | 17 | 10,45 | 1 | 4 |
| 2 | 39,7 | 13,5 | 2292 | 9,241 | 32,76 | 0,97 | 3,6 | 2,926 | 768,1 | 5,205 | 9,654 | 7,133 | 15,8 | 10,07 | 2 | 2 |
| 3 | 38 | 13,2 | 1610 | 9,553 | 33,29 | 0,54 | 2,25 | 3,033 | 462,7 | 5,567 | 12,89 | 11,62 | 15,9 | 9,065 | 2 | 2 |
| 4 | 56,7 | 14,6 | 5548 | 11,3 | 26,41 | 0,76 | 1,48 | 7,899 | 869,2 | 1,682 | 8,192 | 1,98 | 19,3 | 9,8 | 1 | 4 |
| 5 | 53,5 | 10,8 | 2528 | 12,82 | 26,75 | 0,94 | 2,4 | 5,618 | 414,9 | 3,401 | 6,778 | 3,332 | 16,8 | 10,18 | 1 | 4 |
| 6 | 43,4 | 7,5 | 1290 | 9,994 | 25,61 | 0,52 | 2,82 | 4,76 | 216,4 | 2,71 | 5,987 | 1,528 | 15,3 | 7,803 | 2 | 3 |
| … | … | … | … | … | … | … | … | … | … | … | … | … | … | … | … | … |
| 1911 | 55,9 | 15 | 2181 | 14,17 | 39,69 | 0,92 | 0,57 | 3,056 | 343,3 | 3,375 | 3,371 | 1,151 | 11,4 | 0 | 2 | 3 |
| 1912 | 35,7 | 11,9 | 1053 | 11,91 | 40,38 | 0,39 | 1,15 | 2,604 | 146,8 | 1,741 | 3,029 | 0,692 | 10,6 | 0 | 3 | 3 |

Table 4-b: extract from comparison among EURIS classification and Factorial K-means output

In table 5 we present centroids' scores of 4 clusters emerged by Factorial–Kmeans.

|           | Dim.1   | Dim.2   |
|-----------|---------|---------|
| Cluster 1 | -0.6229 | 1.7000  |
| Cluster 2 | 0.3154  | 0.0069  |
| Cluster 3 | -0.2612 | -0.9912 |
| Cluster 4 | 1.4111  | -2.4441 |

Table 5: scores of cluster centroids

In table 6 we present centroids' scores suggestions, as emerged by the meaning of first and second principal components as stated above, and by the relative centroids scores of table 5.

| <<output_ cluster from ClusPca package" result>> | Centroids' scores suggestions |
|---|---|
| 4 | Maximum level of regional R&D expenditures and minimum level of technical-scientific "lag" of individuals belonging to region |
| 2 | Medium-Maximum level of regional R&D expenditures and medium-Maximum level of technical-scientific "lag" of individuals belonging to region |
| 3 | medium-minimum level of regional R&D expenditures and medium-minimum level of technical-scientific "lag" of individuals belonging to region |
| 1 | Minimum level of regional R&D expenditures and maximum level of technical-scientific "lag" of individuals belonging to region |

Table 6: centroids' scores suggestions

By analyzing results of table 6, and by taking in consideration label names of European Union Regional innovation Scoreboard, we stated that: i) centroid's scores of cluster number 4 suggest that this one could be named as <<Innovation leader>> cluster; ii) centroid's scores of cluster number 1 suggest that this one could be named as <<Emerging innovator>> cluster; iii) centroid's scores of cluster number 2 and 3 highlight the need to calculate the euclidean distance of cluster 2 and 3 – with respect to cluster number 4 – in order to define their ranking position in a global innovation ranking among 4 clusters emerged by our Factorial K-means methodology.

After developing the Euclidean distance calculation of cluster 2 and 3 – with respect to cluster number 4 – we present in table 7 the global innovation ranking among 4 clusters defined by our Factorial K-means results; accordingly to results of table 7, we assign to our Factorial K-means cluster number 3 the << Strong innovator>> label, and to our Factorial K-means cluster number 2 the << Moderate innovator>> label.

| "output_ cluster from ClusPca package" result | Innovation ranking | Label assignment of Factorial KMeans |
|---|---|---|
| 4 | First position | Innovation leader |
| 3 | Second position | Strong innovator |
| 2 | Third position | Moderate Innovator |
| 1 | Fourth position | Emerging innovator |

Table 7: label assignment from centroids' scores suggestions

Factorial K-means generates more compact groups than EURIS methodology, resulting in regions having maximum similarities, as in FKM the jointly dimension reduction and cluster analysis problem is addressed by a cluster allocation that minimizes the "within" variance of the clusters in the reduced space (Markos et al., 2019). As a result, our factorial clustering approach is able to present very compact cluster to policymakers, and can push them towards the definition of very targeted policy decisions:

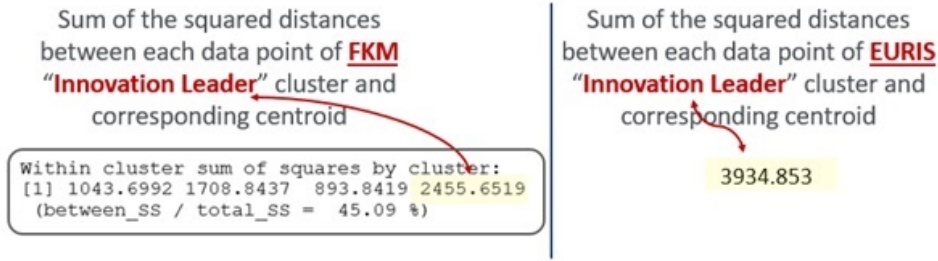

As first analysis of our Factorial K-Means results, We have matched cluster assignment of our FKM methodology with the EURIS one, in order to analyze correlations. With respect to this aim, we noted that cluster number 4 of Factorial K-means (named "Innovation leader" by our label assignment criterion of table 7) is the one that most coincides (73.95%) with "Innovation leader" of EURIS. Cluster number 3 of the factorial K-means (named "Strong innovator" by our label assignment criterion of table 7) coincides at 53.75% with "Strong innovator" of EURIS. Cluster number 2 of the factorial K-means (named "Moderate innovator" by our label assignment criterion of table 7) coincides at 31.16% with "Moderate innovator" of EURIS. Cluster number 1 of the factorial K-means (named "Emerging innovator" by our label assignment criterion of table 7) coincides at 85.35% with "Emerging innovator" of EURIS.

According to above results, we highlight that Factorial K-means methodology has been able to identify correlations between our proposed clusters and the EURIS ones with higher and lower performance levels ("Innovation leader" and "Emerging innovators"); however, it also reveals a significant disparity at the central level ("Strong Innovator" and "Moderate Innovator").

In table 8 we present regions closest to centroids of relative belonging cluster, for each cluster:

| Region | V1 coordinates | V2 coordinates | EUCLIDEAN DISTANCE CALCULATION WITH RESPECT TO RELATIVE CENTROID COORDINATES of Table 5 | Factorial k-means cluster |
|---|---|---|---|---|
| DE3 - Berlin_2023 | 1.629911 | -1.935844 | 0.553362 | 4 (Innovation leader) |
| NO07 - Nord-Norge_2016 | -0.280986 | -1.06179 | 0.07332 | 3 (Strong Innovator) |
| DE27 - Schwaben_2020 | 0.312302 | -0.03747 | 0.044506 | 2 (Moderate Innovator) |
| ITF5 - Basilicata_2016 | -0.636459 | 1.703864 | 0.014094 | 1 (Emerging innovator) |

Table 8: regions closest to centroids of belonging cluster

As stated in previous sections, we use the neural network (NN) methodology for a what-if analysis aimed at predicting the effect of potential regional innovation policies, with the final aim to understand what is the optimal policy path to be addressed - by each region – in order to push the analyzed region into a stronger innovation cluster. By means of NN we are able to understand if specific regional policies (i.e.: specific set of regional innovation indicators values) can give positive contribution to the improvement of regional innovation capability, in order to provide regional policymakers with the opportunity to predict the most effective policy path for promoting regional innovation.

With respect to this general aim and, in particular, with respect to the idea of discovering promising innovation policy paths for a case study (Campania region, Italy), we have developed a Neural Network algorithm for binary classifications of regions; the binary classification algorithm let us to understand which policies should be implemented in order to push the analyzed region into an upper cluster; accordingly, we assigned a binary label (with respect to the "belonging" - or "non belonging" - to a specific cluster) to each region, in order to predict the cluster assignment by means of a what-if analysis.

We split the dataset of table 4-b (1912 observations; 14 features) in training set (1146 data), validation set (382 data) and test set (384 data), and we trained our NN according to the framework of Table 9:

| Binary label: | 1 = cluster 4 (Innovation leader) "belonging"; 0 = cluster 4 (Innovation leader) "non belonging" |
|---|---|
| Scaling of the data: | MinMax procedure between -1 and + 1 (the largest occurring data point corresponds to the maximum value, and the smallest one corresponds to the minimum value) |
| Resampling method: | Undersampling |
| NN Architecture: | ______________________________________________<br>Layer (type)        Output Shape         Param #<br>================================================<br>conv1 (Conv2D)     (None, 3, 16, 32)     320<br><br>flat1 (Flatten)    (None, 1536)          0<br><br>dense (Dense)      (None, 100)           153700<br><br>dense2 (Dense)     (None, 1)             101 |
| Features within the Keras Tensorflow framework: | Conv2D(32,(3,3), padding = "same",activation = 'relu')<br>Dense(100,activation = 'relu')<br>Dense(1,activation = 'sigmoid')<br>keras.optimizers.Adam(learning_rate = 0.01)<br>epochs = 200<br>batch_size = 32 |
| Threshold for classification | 0,50 (i.e.: if predicted probability > 0,50 then predicted class = 1; otherwise, predicted class = 0) |

Table 9: framework of our developed NN

In table 10 we present test set results (precision, recall, accuracy).

| Classification report | | precision | recall |
|---|---|---|---|
| | 0 | 0.98 | 0.99 |
| | 1 | 0.93 | 0.87 |
| Accuracy | | 0.976 | |

Table 10: test set results

As highlighted by Table 10, results are very good: the precision with respect to class 1 is very high (0,93), with a very good relative recall (0,87). Thus, by means of our NN we are able to perform a very reliable what-if analysis.

As case study of a what-if analysis, we have selected Campania Region; this Italian region has been assigned - by our Factorial K-means results - in cluster 1 (Emerging innovator) for years 2016, 2017, 2018, 2019, 2020, 2021, and in cluster 3 (Strong innovator) for years 2022, 2023.

In order to provide Campania's regional policymakers with the opportunity to know the most effective policy path for promoting regional innovation, that is to push the Campania region in Factorial K-means cluster 4 (i.e.: <<Innovation leader>> cluster), we performed a what-if analysis by starting from the 2023 Campania's condition. Thus, we progressively changed innovation indicators of Campania 2023 in order to evaluate which policies should be implemented to promote Campania into the <<Innovation leader>> cluster; more in depth, in order to develop this task we defined the magnitude change of the innovation indicators by simulating innovation indicators values of regions belonging to the Innovation Leader cluster.

In table 11 we present results related to the what-if analysis that has been performed on the base of the above starting point (i.e. on the base of the data related to innovation indicators of Campania 2023)

| Progressive policy trial number | Progressive changes in innovation indicator for Campania region | Predicted probability for the <<belonging>> of Campania to <<Innovation Leader>> cluster of the Factorial K_means |
|---|---|---|
| 1 | First change: 2.2.1 R&D expenditure in the business sector (from 2023 value of CAMPANIA: 0.63 - to 2023 value of HANBURG: 1.22) | 0.00027 % |
| 2 | Progressive further change: 2.1.1 R&D expenditure in the public sector (from 2023 value of CAMPANIA: 0.68 - to 2023 value of HANBURG: 1.04) | 0.00025 % |
| 3 | Progressive further change: 4.1.1 Employment in knowledge-intensive activities (from 2023 value of CAMPANIA: 13 - to 2023 value of HANBURG: 21.8) | 0.028 % |
| 4 | Progressive further change: 2.3.2 Employed ICT specialists (from 2023 value of CAMPANIA: 3.03 - to 2023 value of HANBURG: 8.53) | 38.75 % |

Table 11: progressive what-if analysis for Campania Region

As trial 4 showed the best increase for the probability of <<belonging>> to the Innovation Leader cluster, we further increased - for trial 5 - the indicator <<2.3.2 Employed ICT specialists>> (to BERLIN 2023 value: 11.8). As consequence of trial 5 change, the predicted unitary probability of Campania - for the <<belonging>> to the <<Innovation leader>> cluster – becomes 97.67 %.

With the aim to check a possible overtime datashift issue related to the data distribution of the original EURIS variables, we have run two-sample Kolmogorov-Smirnov (KS) test; the above test is a non-parametric method used to compare the empirical cumulative distribution functions (ECDFs) of two samples to determine if they are drawn from the same underlying distribution.

Let X and Y be two variables, the null and alternative hypotheses of the two-sample Kolmogorov-Smirnov (KS) test are the following one:
H0 : the distribution of X is equal to the distribution of Y.
H1 : the distribution of X is different to the distribution of Y.
When the p-value of the two-sample Kolmogorov-Smirnov test is greater than 0.05, as our selected significance level, there is no evidence to reject the null hypothesis (H0). This means that there is not enough statistical evidence to conclude that the distributions of the two samples (x and y) are significantly different from each other. Thus, according to the idea of testing a possible overtime dataset shift with respect to the original EURIS data, we have split - for each variable - the relative data into three numeric vectors of values:

x for data belonging to 2023, 2022, 2021 years.

y for data belonging to 2020, 2019, 2018 years.

z for data belonging to 2017, 2016 years.

KS results highlight the presence of the overtime datashift issue of above; for example, with respect to "x" and "z" vectors for "2.2.1 R&D expenditure in the business sector", the two-sample Kolmogorov-Smirnov test results are:

D = 0.10112, p-value = 0.005676; alternative hypothesis: two-sided.

According to the above p-value result, for the variable "2.2.1 R&D expenditure in the business sector", the data belonging to the period 2017, 2016 do not have the same underlying distribution of the data belonging to the period 2023, 2022, 2021. Thus, for this variable there has been a distribution dataset shift between the above-analyzed two data periods.

Empirical cumulative distribution function (ECDF)s - with respect to "x" and "z" periods for the variable "2.2.1 R&D expenditure in the business sector" - is presented in fig. 4-a

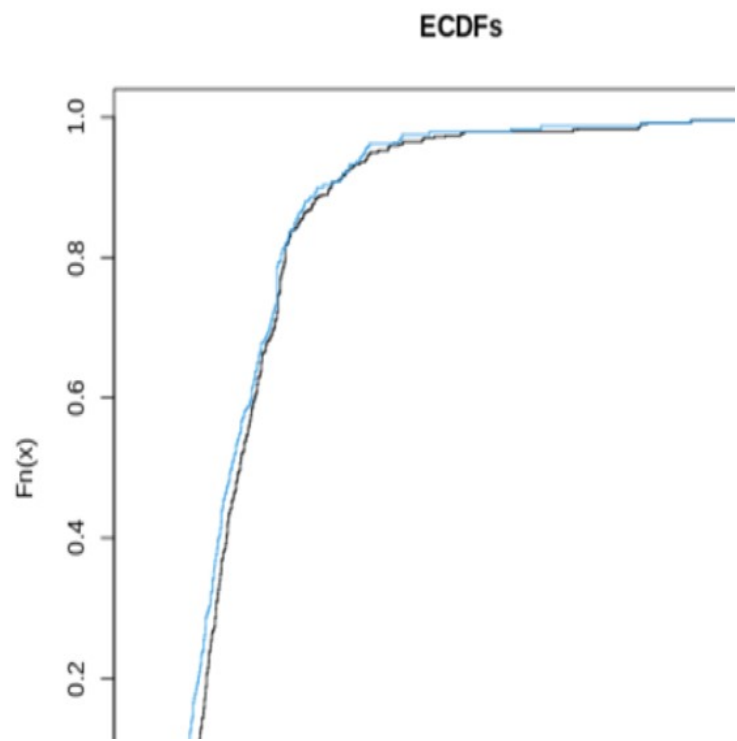

Fig. 4-a - KS results: empirical cumulative distribution function, with respect to "x" and "z" periods, of "2.2.1 R&D expenditure in the business sector"

We also highlight that other EURIS original variables do not suffer from the overtime data shift issue of above; for example, with respect to "x" and "z" vectors for "2.1.1 R&D expenditure in the public sector', the results of the two sample Kolmogorov-Smirnov test are:

D = 0.03696, p-value = 0.8282; alternative hypothesis: two-sided

Empirical cumulative distribution function - with respect to "x" and "z" periods for the variable "2.1.1 R&D expenditure in the public sector" – is presented in fig. 4-b

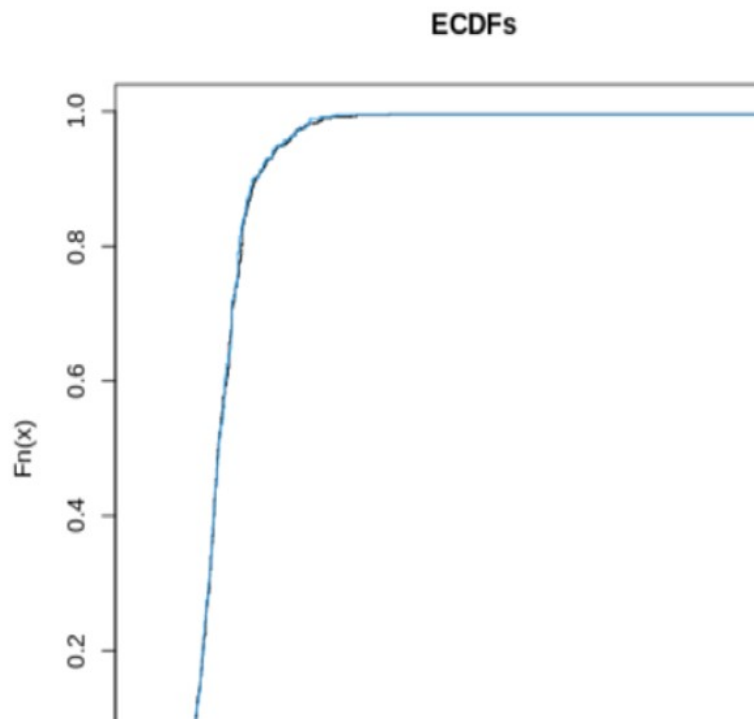

Fig. 4-b - KS results: empirical cumulative distribution function, with respect to "x" and "z" periods, of "2.1.1 R&D expenditure in the public sector"

As consequence of the above hypothesis test results, this work also highlights that the presence of the overtime distribution shift within the EURIS original data; moreover, it points out that the overtime distribution shift is able to have - generally speaking - a relevant impact on the development of what-if analysis methodologies; indeed, dataset shift can lead to significant deterioration in the performance of machine learning systems, if not well addressed with appropriate transfer learning techniques that are able to reuse the trained model knowledge for learning a new task. From this last point of view, we highlight that our proposed specific machine learning solution for what-if analysis, as neural network-based solution, is able to represents a very useful machine learning tool because it is able to address the overtime distribution shift issue of above, as the Transfer Learning technique can be implemented within a Neural Network-based framework in a very effective way.

As highlighted in section 2.1, it is interesting to compare different grouping/labeling methodologies; from last point of view, generally speaking we highlight that clustering methods based on unsupervised techniques are weaker than those based on supervised techniques, according to the absence of ground truth observations; in order to have the opportunity to use supervised clustering techniques even in current case where the original data do not have ground truth observations, we identify a set of pivot regions that are capable of representing a sort of "ground truth regions" for a supervised clustering technique; in particular, we consider pivot regions the ones that belong to the distance range from 0 to 1 standard deviation with respect to the centroids identified by our adopted unsupervised clustering technique (i.e.: FKM); thus, in accordance with Aparicio- Navarro (2014) we group all other regions (i.e.: the "non Pivot" regions) by making their relative assignments with the supervised binary classification algorithm (i.e., by evaluation of membership, or non-membership, in the i-th group, constituted by only pivot regions, identified with our FKM methodology) of table 9; the relative final labeling, then, is developed in consideration of the labels assigned - to the pivot regions - in accordance with the relative EURIS labels. This additional clustering technique, therefore, allows fine tuning of the labeling assignments which arise from the Factorial KMeans methodology. More in depth on this point, according to Aparicio-Navarro et al. (2014) we adopt a methodology - which is based on the analysis of the Euclidean distance distribution among each region (belonging to a specific cluster) and its relative centroid - by using mean and standard deviation of the above distance: accordingly, we assign the most closed regions of each relative centroid to each cluster, and – as consequence - we fine tune the assigned FKM labels by excluding the "border" regions (i.e.: the regions with euclidean distance - with respect to the relative centroid - greater than 1 standard deviation). This fine-tuning process of FKM results - according to Aparicio-Navarro et al. (2014) idea to exclude "border" regions - let us to develop 4 fine-tuned clusters characterized by only the presence of pivot regions within each cluster. Results of the above fine tuning process, related to Pivot regions percentage of every cluster - on total number of regions assigned by FKM - are listed below:

- Pivot regions percentage of cluster 4 ('Innovation leader') on total number of regions assigned by FKM : 91 %

- Pivot regions percentage of cluster 3 ('Strong Innovator') on total number of regions assigned by FKM : 70 %

- Pivot regions percentage of cluster 2 ('Moderate Innovator') on total number of regions assigned by FKM : 61 %

- Pivot regions percentage of cluster 1 ('Emerging Innovator') on total number of regions assigned by FKM : 68 %

As pivot regions are the ones with euclidean distance – with respect to the relative centroid of the FKM clustering method - belonging to the range from 0 to 1 standard deviation, each

FKM cluster developed under the "Aparicio-Navarro et. al (2014) idea" has to be considered a "fine-tuned" FKM cluster; indeed, each "fine-tuned" FKM cluster is necessarily characterized by a lower "deviance within" than the corresponding "non fine-tuned" FKM cluster. According to literature idea that clustering methods based on unsupervised techniques - such as FKM for example - are weaker than those based on supervised techniques, we highlight - for the regional innovation capability topic - the importance to exploit the "Aparicio-Navarro et. al (2014) idea" in order to also create groups under a supervised grouping/labeling process conducted by a binary classification algorithm (that - in our case - can be directly used as supervised grouping technique, on EURIS data, by means of the exploitation of the above pivot regions as a sort of ground truth regions).

With the aim to compare clusters "coherence" on each labeled dataset developed by each one of the above three grouping methodologies, we have trained the same binary classification algorithm (i.e.: a convolutional neural network, according to architecture presented in table 9 of current section) in order to analyze relative classification reports.

Comparison among test set results of three experimented dataset (EURIS "innovation leader" original group Vs FKM "innovation leader" group Vs fine-tuned FKM "innovation leader" group), such as specific outputs of the above three different grouping methodologies, gives insights on the relative internal "coherence" - for each one of the above groups - implicitly recognized by the classification algoritm.

Furthermore, we have also trained the same binary classification algorithm - of above - on an "intersection dataset" between EURIS labeled dataset and FKM labeled dataset.

CNN classification reports - for each dataset developed by each one of the above three different grouping methodologies (with respect to "Innovation leader" cluster) and for the above "intersection dataset" - are presented below:

Accuracy, precision and recall of the binary classification algorithm with respect to "innovation leader" group of the EURIS grouping/labeling methodology, on 14 original EURIS variables values:

|   | precision | recall |
|---|---|---|
| 0 | 0.96 | 0.98 |
| 1 | 0.85 | 0.77 |

Accuracy: 0.950

Note: class 1 = regions belonging to "innovation leader" group; class 0 = regions non belonging to "innovation Leader" group

****************

Accuracy, precision and recall of the binary classification algorithm with respect to "innovation leader" group of the FKM grouping/labeling methodology, on 14 original EURIS variables values:

|   | precision | recall |
|---|---|---|
| 0 | 0.98 | 0.99 |
| 1 | 0.93 | 0.87 |

Accuracy: 0.976

Note: class 1 = regions belonging to "innovation leader" group; class 0 = regions non belonging to "innovation Leader" group

****************

Accuracy, precision and recall of the binary classification algorithm with respect to "innovation leader" group of the fine-tuned FKM grouping/labeling methodology, on 14 original EURIS variables values:

|   | precision | recall |
|---|---|---|
| 0 | 0.99 | 0.98 |
| 1 | 0.87 | 0.93 |

Accuracy: 0.976

Note: class 1 = regions belonging to "innovation leader" group; class 0 = regions non belonging to "innovation Leader" group

****************

Accuracy, precision and recall of the binary classification algorithm with respect to "innovation leader" group of the "intersection" dataset between EURIS grouping/labeling methodology and FKM grouping/labeling methodology, on 14 original EURIS variables values:

|   | precision | recall |
|---|---|---|
| 0 | 0.99 | 0.99 |
| 1 | 0.95 | 0.95 |

Note: class 1 = regions belonging to "innovation leader" group; class 0 = regions non belonging to "innovation Leader" group

Accuracy: 0.989

Experimental results of the above comparison among different grouping/labelling methodologies suggest that each one of our proposed grouping/labelling methods (i.e., FKM methodology and "fine tuned" FKM one) are able to develop more cohesive groups, resulting in regions having better similarities, than the ones developed by the EURIS methodology.

# 5 Discussion

According to results of 4.4 section, the Factorial K-means methodology has been able to identify correlations, between our clusters and the EURIS ones, with higher and lower performance levels ("Innovation leader" and "Emerging innovators"); however, it also revealed a significant disparity at the central level ("Strong Innovator" and "Moderate Innovator"); such variation has implications for cluster assignment and is critical for informing policy decisions.

Factorial K-means generates more compact groups than EURIS methodology, resulting in regions having maximum similarities, as in FKM the jointly dimension reduction and cluster analysis problem is addressed by a cluster allocation that minimizes the "within" variance of the clusters in the reduced space (Markos et al., 2019). As a result, our factorial clustering approach is able to present very compact cluster to policymakers, and can push them towards the definition of very targeted policy decisions. Moreover, the factorial clustering approach is also critical to crafting policies that are also fair from the point of view of their impact. Such precision and fairness in policymaking adds substantial value to the usefulness of the research conducted, and let us to state that our proposed FKM-NN tool could be used as an effective synergistic instruments of the European Regional Innovation Scoreboard.

Currently, policymakers rely heavily on Regional Innovation Scoreboards to make decisions about policy formulation and implementation, with the aim to allocate resources and adapt policies to address specific weaknesses and to improve overall regional innovation capabilities in a targeted way (Teirlinck & Spithoven, 2023); but we claim that there are analysis examples – also - that are able to get further stronger the need to develop a synergistic tool of European regional Innovation Scoreobard for the definition of optimal innovation policies: Andalusia region (Spain) is an illustrative case study on this point.

"S4Andalucia" is the regional agency of the European Cohesion Framework 2021-2027 which includes the planning, execution, development and evaluation of public policies, ranging from the field of research to that of entrepreneurship; this agency used EURIS innovation indicators to monitor the region's performance, with the aim to identify areas to be improved, and with the objective to develop policies aimed at increasing the region's innovative capacity; according to EURIS results, Andalusia has been always classified as a "Moderate Innovator" region for every year from 2016 to 2023; thus, according to EURIS results, the situation in Andalusia remained unchanged during the whole period of the EURIS analysis. The above unchanged situation could indicate that policy interventions have not been effective, or that the EURIS "Moderate Innovator" classification has not been appropriate in all the years.

By using our Factorial K-means approach, we state that Andalusia is included in <<Emerging innovator>> cluster in 2016, 2017, 2018, 2019, 2020, and in <<Strong Innovator>> one in 2021, 2022, 2023; thus, according to our cohesive clustering approach, we highlight that Andalusia region has improved its innovation capability overtime (from <<emerging>> to <<strong>>); in order to exploit the discovery efforts of FKM methodology – with respect to our very cohesive clusters assignment - we also claim the necessity of using new methodologies, such as the neural network one of 4.4 section, to define very targeted policy interventions; more on this point, according to 4.4 section results of Campania case study, we also claim that a NN-based what-if analysis tool is able to show which policies should be implemented in order to push an analyzed region into an upper cluster, in coherence with the idea that a specific innovation policy has to be considered potentially effective if it is able to strongly increase the predicted probability - of the analyzed region - to belong to an upper class; as example, we observed in 4.4 section that the most effective policy for Campania region, to belong to the "Innovation leader" cluster, should be aimed at increasing the value of "2.3.2 Employed ICT specialists".

# 6 Conclusions

The European Regional Innovation Scoreboard is currently and broadly used for the definition of regional innovation policies by European policymakers; it is a regional innovation measuring tool for the over time change analysis of each specific innovation indicator, from which it is possible to analyze the evolution of each regional innovation indicator; according to the importance of the European Regional Innovation Scoreboard for innovation policy purposes, we state that European regional policymakers need integrative and synergistic methodological tools, of EURIS, for innovation policy purposes. More in depth, we highlight the need to integrate the current methodology of the European Regional Innovation Scoreboard with a Factorial K-means tool for clustering purposes, and with a neural network tool for performing what-if policy analyses.

Factorial K-means generates more compact groups than EURIS methodology, resulting in regions having maximum similarities, as in FKM the jointly dimension reduction and cluster analysis problem is addressed by a cluster allocation that minimizes the "within" variance of the clusters in the reduced space. As a result, our factorial clustering approach is able to present very cohesive cluster to policymakers, and can push them towards the definition of very targeted policy decisions. Moreover, the factorial clustering approach is also critical to crafting policies that are also fair from the point of view of their impact. Such precision and fairness in policymaking adds substantial value to the usefulness of the research conducted; thus, our proposed FKM-NN tool could be used as an effective synergistic instruments of the European Regional Innovation Scoreboard.

Experimental results of the comparison among different grouping/labelling methodologies suggest that each one of our proposed grouping/labelling methodologies (i.e., FKM methodology and "fine tuned" FKM one) are able to develop more cohesive groups, resulting in regions having better similarities, than the ones developed by the EURIS methodology.

As main conclusion of this study, we claim the necessity of using our methodology as a synergistic tool of the European Regional innovation Scoreboard, in order to help policymakers toward the definition of very targeted innovation policies that are aimed at exploiting the discovery efforts for very cohesive FKM clusters assignment; more on this point, we also claim that a neural network-based what-if analysis tool is able to show which policies should be implemented in order to push an analyzed region into a better cluster, from the innovation level point of view; moreover, we state that a neural network methodology is also able to highlight what is the effectiveness of each possible and specific innovation policy, according to the idea that a specific innovation policy has to be considered potentially effective if it is able to strongly increase the predicted probability - of the analyzed region - to belong to a better innovation cluster.

Furthermore, by analyzing the data of the variables underlying the development of regional innovative capacity, a critical aspect emerged - in the context of machine learning approaches - related to the problem of "dataset shift"; this problem occurs with probability distributions that - within the same variable – can change "overtime". Literature points out that dataset shift can lead to significant deterioration in the performance of ML systems, if not well addressed with appropriate transfer learning techniques; thus, we highlight the need to take into account the possible presence of the overtime distribution shift, for to the development of effective what-if analysis methodologies. From this last point of view, we claim that our proposed specific machine learning solution for what-if analysis, as neural network-based solution, is able to represents a very useful machine learning tool because it is able to address the overtime distribution shift issue of above, as the Transfer Learning technique can be implemented within a Neural Network-based framework in a very effective way.

## Website references

# Credit authorship contribution statement

# Acknowledgments

The support of Antonio Impiombato and Sara Ianniello is gratefully acknowledged

# Annex 1: Correlation matrix and PCA results

Correlation matrix

```
# Correlation Matrix (auto-method)

Parameter1                                              |                                                Parameter2 |        r |         95% CI | t(1910) |        p
---------------------------------------------------------------------------------------------------------------------------------------------------------------------
1.1.2 Population with tertiary education                |                   1.1.3 Population involved in lifelong learning |     0.48 | [ 0.45,  0.52] |   24.00 | < .001***
1.1.2 Population with tertiary education                |                    1.2.1 International scientific co-publications |     0.56 | [ 0.53,  0.59] |   29.74 | < .001***
1.1.2 Population with tertiary education                |       1.2.2 Scientific publications among the top 10% most cited |     0.38 | [ 0.34,  0.42] |   18.08 | < .001***
1.1.2 Population with tertiary education                |          1.3.2 Individuals with above basic overall digital skills |     0.60 | [ 0.57,  0.63] |   32.94 | < .001***
1.1.2 Population with tertiary education                |                        2.1.1 R&D expenditure in the public sector |     0.23 | [ 0.18,  0.27] |   10.20 | < .001***
1.1.2 Population with tertiary education                |                      2.2.1 R&D expenditure in the business sector |     0.26 | [ 0.22,  0.30] |   11.93 | < .001***
1.1.2 Population with tertiary education                |                                     2.3.2 Employed ICT specialists |     0.61 | [ 0.58,  0.63] |   33.38 | < .001***
1.1.2 Population with tertiary education                |                                 3.2.2 Public-private co-publications |     0.46 | [ 0.42,  0.49] |   22.35 | < .001***
1.1.2 Population with tertiary education                |                                       3.3.1 PCT patent applications |     0.25 | [ 0.21,  0.29] |   11.21 | < .001***
1.1.2 Population with tertiary education                |                                         3.3.2 Trademark applications |     0.36 | [ 0.32,  0.40] |   16.76 | < .001***
1.1.2 Population with tertiary education                |                                            3.3.3 Design applications | 7.83e-03 | [-0.04,  0.05] |    0.34 |     0.898
1.1.2 Population with tertiary education                |                 4.1.1 Employment in knowledge-intensive activities |     0.36 | [ 0.32,  0.39] |   16.67 | < .001***
1.1.2 Population with tertiary education                |                           4.3.2 Air emissions by fine particulates |    -0.21 | [-0.25, -0.16] |   -9.21 | < .001***
1.1.3 Population involved in lifelong learning          |                    1.2.1 International scientific co-publications |     0.56 | [ 0.53,  0.59] |   29.83 | < .001***
1.1.3 Population involved in lifelong learning          |       1.2.2 Scientific publications among the top 10% most cited |     0.56 | [ 0.53,  0.59] |   29.87 | < .001***
1.1.3 Population involved in lifelong learning          |          1.3.2 Individuals with above basic overall digital skills |     0.76 | [ 0.74,  0.78] |   50.71 | < .001***
1.1.3 Population involved in lifelong learning          |                        2.1.1 R&D expenditure in the public sector |     0.15 | [ 0.11,  0.19] |    6.69 | < .001***
1.1.3 Population involved in lifelong learning          |                      2.2.1 R&D expenditure in the business sector |     0.27 | [ 0.22,  0.31] |   12.09 | < .001***
1.1.3 Population involved in lifelong learning          |                                     2.3.2 Employed ICT specialists |     0.45 | [ 0.41,  0.49] |   22.04 | < .001***
1.1.3 Population involved in lifelong learning          |                                 3.2.2 Public-private co-publications |     0.50 | [ 0.47,  0.54] |   25.54 | < .001***
1.1.3 Population involved in lifelong learning          |                                       3.3.1 PCT patent applications |     0.48 | [ 0.44,  0.51] |   23.73 | < .001***
1.1.3 Population involved in lifelong learning          |                                         3.3.2 Trademark applications |     0.38 | [ 0.35,  0.42] |   18.19 | < .001***
1.1.3 Population involved in lifelong learning          |                                            3.3.3 Design applications |     0.07 | [ 0.03,  0.12] |    3.20 |    0.013*
1.1.3 Population involved in lifelong learning          |                 4.1.1 Employment in knowledge-intensive activities |     0.20 | [ 0.16,  0.25] |    9.05 | < .001***
1.1.3 Population involved in lifelong learning          |                           4.3.2 Air emissions by fine particulates |    -0.52 | [-0.55, -0.48] |  -26.47 | < .001***
1.2.1 International scientific co-publications          |       1.2.2 Scientific publications among the top 10% most cited |     0.56 | [ 0.53,  0.59] |   29.42 | < .001***
1.2.1 International scientific co-publications          |          1.3.2 Individuals with above basic overall digital skills |     0.42 | [ 0.39,  0.46] |   20.39 | < .001***
1.2.1 International scientific co-publications          |                        2.1.1 R&D expenditure in the public sector |     0.40 | [ 0.36,  0.44] |   19.19 | < .001***
1.2.1 International scientific co-publications          |                      2.2.1 R&D expenditure in the business sector |     0.38 | [ 0.34,  0.42] |   17.86 | < .001***
1.2.1 International scientific co-publications          |                                     2.3.2 Employed ICT specialists |     0.60 | [ 0.57,  0.63] |   33.00 | < .001***
1.2.1 International scientific co-publications          |                                 3.2.2 Public-private co-publications |     0.91 | [ 0.90,  0.92] |   96.28 | < .001***
1.2.1 International scientific co-publications          |                                       3.3.1 PCT patent applications |     0.43 | [ 0.39,  0.46] |   20.63 | < .001***
1.2.1 International scientific co-publications          |                                         3.3.2 Trademark applications |     0.37 | [ 0.34,  0.41] |   17.64 | < .001***
1.2.1 International scientific co-publications          |                                            3.3.3 Design applications |     0.02 | [-0.02,  0.07] |    0.90 |     0.898
1.2.1 International scientific co-publications          |                 4.1.1 Employment in knowledge-intensive activities |     0.39 | [ 0.36,  0.43] |   18.77 | < .001***
1.2.1 International scientific co-publications          |                           4.3.2 Air emissions by fine particulates |    -0.30 | [-0.34, -0.25] |  -13.55 | < .001***
1.2.2 Scientific publications among the top 10% most cited |          1.3.2 Individuals with above basic overall digital skills |     0.57 | [ 0.53,  0.60] |   29.98 | < .001***
1.2.2 Scientific publications among the top 10% most cited |                        2.1.1 R&D expenditure in the public sector |     0.16 | [ 0.11,  0.20] |    7.01 | < .001***
1.2.2 Scientific publications among the top 10% most cited |                      2.2.1 R&D expenditure in the business sector |     0.26 | [ 0.22,  0.30] |   11.90 | < .001***
1.2.2 Scientific publications among the top 10% most cited |                                     2.3.2 Employed ICT specialists |     0.35 | [ 0.31,  0.39] |   16.43 | < .001***
1.2.2 Scientific publications among the top 10% most cited |                                 3.2.2 Public-private co-publications |     0.54 | [ 0.51,  0.57] |   27.92 | < .001***
1.2.2 Scientific publications among the top 10% most cited |                                       3.3.1 PCT patent applications |     0.51 | [ 0.48,  0.55] |   26.10 | < .001***
1.2.2 Scientific publications among the top 10% most cited |                                         3.3.2 Trademark applications |     0.35 | [ 0.30,  0.38] |   16.07 | < .001***
1.2.2 Scientific publications among the top 10% most cited |                                            3.3.3 Design applications |     0.13 | [ 0.08,  0.17] |    5.62 | < .001***
1.2.2 Scientific publications among the top 10% most cited |                 4.1.1 Employment in knowledge-intensive activities |     0.22 | [ 0.18,  0.27] |   10.09 | < .001***
1.2.2 Scientific publications among the top 10% most cited |                           4.3.2 Air emissions by fine particulates |    -0.52 | [-0.55, -0.48] |  -26.34 | < .001***
1.3.2 Individuals with above basic overall digital skills |                        2.1.1 R&D expenditure in the public sector |     0.12 | [ 0.08,  0.17] |    5.34 | < .001***
1.3.2 Individuals with above basic overall digital skills |                      2.2.1 R&D expenditure in the business sector |     0.20 | [ 0.16,  0.24] |    8.90 | < .001***
1.3.2 Individuals with above basic overall digital skills |                                     2.3.2 Employed ICT specialists |     0.34 | [ 0.30,  0.38] |   15.67 | < .001***
1.3.2 Individuals with above basic overall digital skills |                                 3.2.2 Public-private co-publications |     0.35 | [ 0.31,  0.39] |   16.53 | < .001***
1.3.2 Individuals with above basic overall digital skills |                                       3.3.1 PCT patent applications |     0.26 | [ 0.22,  0.30] |   11.95 | < .001***
1.3.2 Individuals with above basic overall digital skills |                                         3.3.2 Trademark applications |     0.21 | [ 0.17,  0.26] |    9.52 | < .001***
1.3.2 Individuals with above basic overall digital skills |                                            3.3.3 Design applications |    -0.04 | [-0.08,  0.01] |   -1.73 |     0.419
1.3.2 Individuals with above basic overall digital skills |                 4.1.1 Employment in knowledge-intensive activities |     0.07 | [ 0.02,  0.11] |    2.86 |    0.030*
1.3.2 Individuals with above basic overall digital skills |                           4.3.2 Air emissions by fine particulates |    -0.55 | [-0.58, -0.52] |  -28.96 | < .001***
2.1.1 R&D expenditure in the public sector              |                      2.2.1 R&D expenditure in the business sector |     0.53 | [ 0.50,  0.57] |   27.59 | < .001***
2.1.1 R&D expenditure in the public sector              |                                     2.3.2 Employed ICT specialists |     0.24 | [ 0.20,  0.28] |   10.96 | < .001***
2.1.1 R&D expenditure in the public sector              |                                 3.2.2 Public-private co-publications |     0.42 | [ 0.38,  0.45] |   19.94 | < .001***
2.1.1 R&D expenditure in the public sector              |                                       3.3.1 PCT patent applications |     0.19 | [ 0.14,  0.23] |    8.34 | < .001***
2.1.1 R&D expenditure in the public sector              |                                         3.3.2 Trademark applications |     0.11 | [ 0.06,  0.15] |    4.72 | < .001***
2.1.1 R&D expenditure in the public sector              |                                            3.3.3 Design applications |    -0.03 | [-0.07,  0.02] |   -1.21 |     0.898
2.1.1 R&D expenditure in the public sector              |                 4.1.1 Employment in knowledge-intensive activities |     0.19 | [ 0.15,  0.24] |    8.67 | < .001***
2.1.1 R&D expenditure in the public sector              |                           4.3.2 Air emissions by fine particulates |    -0.03 | [-0.07,  0.02] |   -1.22 |     0.898
2.2.1 R&D expenditure in the business sector            |                                     2.3.2 Employed ICT specialists |     0.39 | [ 0.35,  0.42] |   18.34 | < .001***
2.2.1 R&D expenditure in the business sector            |                                 3.2.2 Public-private co-publications |     0.51 | [ 0.48,  0.55] |   26.12 | < .001***
2.2.1 R&D expenditure in the business sector            |                                       3.3.1 PCT patent applications |     0.54 | [ 0.51,  0.57] |   28.11 | < .001***
2.2.1 R&D expenditure in the business sector            |                                         3.3.2 Trademark applications |     0.28 | [ 0.24,  0.32] |   12.80 | < .001***
2.2.1 R&D expenditure in the business sector            |                                            3.3.3 Design applications |     0.16 | [ 0.12,  0.20] |    7.12 | < .001***
2.2.1 R&D expenditure in the business sector            |                 4.1.1 Employment in knowledge-intensive activities |     0.51 | [ 0.48,  0.54] |   26.03 | < .001***
2.2.1 R&D expenditure in the business sector            |                           4.3.2 Air emissions by fine particulates |    -0.12 | [-0.16, -0.07] |   -5.08 | < .001***
2.3.2 Employed ICT specialists                          |                                 3.2.2 Public-private co-publications |     0.55 | [ 0.52,  0.58] |   28.88 | < .001***
2.3.2 Employed ICT specialists                          |                                       3.3.1 PCT patent applications |     0.39 | [ 0.35,  0.42] |   18.31 | < .001***
2.3.2 Employed ICT specialists                          |                                         3.3.2 Trademark applications |     0.48 | [ 0.45,  0.51] |   23.96 | < .001***
2.3.2 Employed ICT specialists                          |                                            3.3.3 Design applications |     0.10 | [ 0.06,  0.15] |    4.44 | < .001***
2.3.2 Employed ICT specialists                          |                 4.1.1 Employment in knowledge-intensive activities |     0.71 | [ 0.69,  0.73] |   44.32 | < .001***
2.3.2 Employed ICT specialists                          |                           4.3.2 Air emissions by fine particulates |    -0.15 | [-0.19, -0.10] |   -6.49 | < .001***
3.2.2 Public-private co-publications                    |                                       3.3.1 PCT patent applications |     0.54 | [ 0.51,  0.57] |   28.22 | < .001***
3.2.2 Public-private co-publications                    |                                         3.3.2 Trademark applications |     0.38 | [ 0.34,  0.41] |   17.80 | < .001***
3.2.2 Public-private co-publications                    |                                            3.3.3 Design applications |     0.07 | [ 0.03,  0.12] |    3.23 |    0.013*
3.2.2 Public-private co-publications                    |                 4.1.1 Employment in knowledge-intensive activities |     0.42 | [ 0.38,  0.46] |   20.15 | < .001***
3.2.2 Public-private co-publications                    |                           4.3.2 Air emissions by fine particulates |    -0.28 | [-0.32, -0.24] |  -12.62 | < .001***
3.3.1 PCT patent applications                           |                                         3.3.2 Trademark applications |     0.43 | [ 0.39,  0.46] |   20.75 | < .001***
3.3.1 PCT patent applications                           |                                            3.3.3 Design applications |     0.36 | [ 0.32,  0.40] |   17.04 | < .001***
3.3.1 PCT patent applications                           |                 4.1.1 Employment in knowledge-intensive activities |     0.49 | [ 0.45,  0.52] |   24.54 | < .001***
3.3.1 PCT patent applications                           |                           4.3.2 Air emissions by fine particulates |    -0.30 | [-0.34, -0.26] |  -13.92 | < .001***
3.3.2 Trademark applications                            |                                            3.3.3 Design applications |     0.46 | [ 0.42,  0.49] |   22.50 | < .001***
3.3.2 Trademark applications                            |                 4.1.1 Employment in knowledge-intensive activities |     0.38 | [ 0.34,  0.42] |   18.05 | < .001***
3.3.2 Trademark applications                            |                           4.3.2 Air emissions by fine particulates |    -0.12 | [-0.17, -0.08] |   -5.48 | < .001***
3.3.3 Design applications                               |                 4.1.1 Employment in knowledge-intensive activities |     0.24 | [ 0.19,  0.28] |   10.61 | < .001***
3.3.3 Design applications                               |                           4.3.2 Air emissions by fine particulates |     0.07 | [ 0.03,  0.12] |    3.11 |    0.015*
4.1.1 Employment in knowledge-intensive activities      |                           4.3.2 Air emissions by fine particulates |     0.06 | [ 0.01,  0.10] |    2.58 |     0.059

p-value adjustment method: Holm (1979)
Observations: 1912
```

-----------------

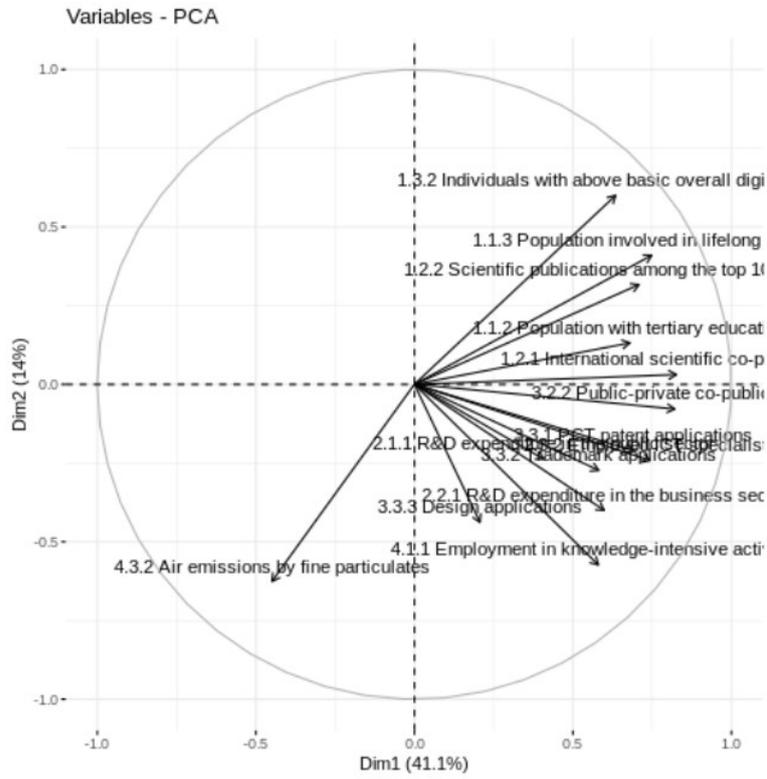

PCA results on standardized data

-----------------

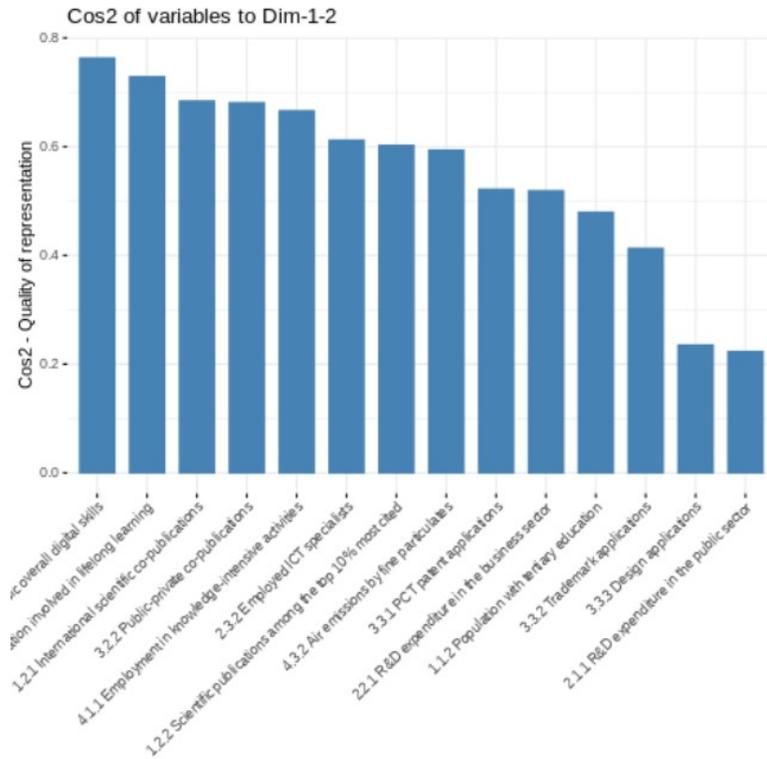

PCA results

-----------------

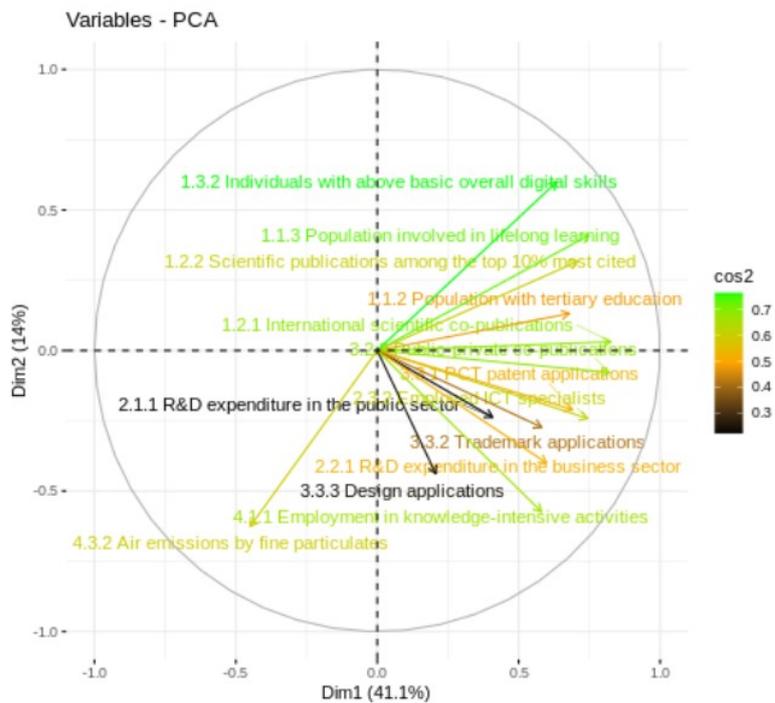

PCA results

----------------

```
                                                              Comp.1
1.1.2 Population with tertiary education                      0.28317975
1.1.3 Population involved in lifelong learning                0.31188749
1.2.1 International scientific co-publications                0.34436626
1.2.2 Scientific publications among the top 10% most cited    0.29519691
1.3.2 Individuals with above basic overall digital skills     0.26442142
2.1.1 R&D expenditure in the public sector                    0.17002323
2.2.1 R&D expenditure in the business sector                  0.24948558
2.3.2 Employed ICT specialists                                0.30999915
3.2.2 Public-private co-publications                          0.34218824
3.3.1 PCT patent applications                                 0.28781480
3.3.2 Trademark applications                                  0.24228041
3.3.3 Design applications                                     0.08624897
4.1.1 Employment in knowledge-intensive activities            0.24184859
4.3.2 Air emissions by fine particulates                     -0.18728962
```

```
                              Comp.2
1.1.2 Population with tertiary education                        0.09408124
1.1.3 Population involved in lifelong learning                  0.29290848
1.2.1 International scientific co-publications                  0.02217086
1.2.2 Scientific publications among the top 10% most cited      0.22640293
1.3.2 Individuals with above basic overall digital skills       0.42860727
2.1.1 R&D expenditure in the public sector                     -0.16992236
2.2.1 R&D expenditure in the business sector                   -0.28590943
2.3.2 Employed ICT specialists                                 -0.17256064
3.2.2 Public-private co-publications                           -0.05598332
3.3.1 PCT patent applications                                  -0.15028056
3.3.2 Trademark applications                                   -0.19518955
3.3.3 Design applications                                      -0.31312128
4.1.1 Employment in knowledge-intensive activities             -0.40971181
4.3.2 Air emissions by fine particulates                       -0.44696565

                              Comp.3
1.1.2 Population with tertiary education                        0.126486015
1.1.3 Population involved in lifelong learning                 -0.124538368
1.2.1 International scientific co-publications                  0.225685872
1.2.2 Scientific publications among the top 10% most cited     -0.155331118
1.3.2 Individuals with above basic overall digital skills      -0.040902808
2.1.1 R&D expenditure in the public sector                      0.486984213
2.2.1 R&D expenditure in the business sector                    0.188100933
2.3.2 Employed ICT specialists                                  0.078659759
3.2.2 Public-private co-publications                            0.201647578
3.3.1 PCT patent applications                                  -0.256746812
3.3.2 Trademark applications                                   -0.380387782
3.3.3 Design applications                                      -0.585793424
4.1.1 Employment in knowledge-intensive activities              0.004034089
4.3.2 Air emissions by fine particulates                        0.136337402

                              Comp.4
1.1.2 Population with tertiary education                        0.46335396
1.1.3 Population involved in lifelong learning                  0.01725515
1.2.1 International scientific co-publications                  0.04186970
1.2.2 Scientific publications among the top 10% most cited     -0.17925134
1.3.2 Individuals with above basic overall digital skills       0.10299975
2.1.1 R&D expenditure in the public sector                     -0.35381901
2.2.1 R&D expenditure in the business sector                   -0.36963706
2.3.2 Employed ICT specialists                                  0.42858836
3.2.2 Public-private co-publications                           -0.12035833
3.3.1 PCT patent applications                                  -0.33841491
3.3.2 Trademark applications                                    0.15447749
3.3.3 Design applications                                      -0.13258469
4.1.1 Employment in knowledge-intensive activities              0.23317204
4.3.2 Air emissions by fine particulates                        0.26492056

                            PCA results

                            ---------------
```

```
        Eigenvalue variance.  Percent cumulative.   Variance.percent
Dim.1   5.76009622            41.1435444                   41.14354
Dim.2   1.96016931            14.0012094                   55.14475
Dim.3   1.40769345            10.0549532                   65.19971
Dim.4   1.11137752             7.9384108                   73.13812
Dim.5   0.74551365             5.3250975                   78.46322
Dim.6   0.72981916             5.2129940                   83.67621
Dim.7   0.47980632             3.4271880                   87.10340
Dim.8   0.42690114             3.0492939                   90.15269
Dim.9   0.37658725             2.6899089                   92.84260
Dim.10  0.35238312             2.5170223                   95.35962
Dim.11  0.28888394             2.0634567                   97.42308
Dim.12  0.16861254             1.2043753                   98.62745
Dim.13  0.13479526             0.9628233                   99.59028
Dim.14  0.05736112             0.4097223                  100.00000
```

                              PCA results
                              -----------

# Annex 2 : Factorial K-means results

```
Cluster centroids:
            Dim.1    Dim.2
Cluster 1 -0.6229   1.7000
Cluster 2  0.3154   0.0069
Cluster 3 -0.2612  -0.9912
Cluster 4  1.4111  -2.4441
```

----------------

```
Variable scores:
                                                          Dim.1    Dim.2
1.1.2 Population with tertiary education                -0.1814  -0.0848
1.1.3 Population involved in lifelong learning          -0.1226  -0.3924
1.2.1 International scientific co-publications           0.1413  -0.2008
1.2.2 Scientific publications among the top 10% most cited  0.0260  -0.4246
1.3.2 Individuals with above basic overall digital skills  -0.1981  -0.3416
2.1.1 R&D expenditure in the public sector               0.6283   0.0470
2.2.1 R&D expenditure in the business sector             0.5583  -0.0758
2.3.2 Employed ICT specialists                          -0.0916  -0.0013
3.2.2 Public-private co-publications                     0.2782  -0.2327
3.3.1 PCT patent applications                            0.2617  -0.3602
3.3.2 Trademark applications                            -0.1005  -0.3239
3.3.3 Design applications                               -0.0125  -0.2420
4.1.1 Employment in knowledge-intensive activities       0.1127   0.0894
4.3.2 Air emissions by fine particulates                 0.0920   0.3718
```

----------------

```
  Clusters size:

Cluster 1 – 679 regions
Cluster 2 – 477 regions
Cluster 3 – 475 regions
Cluster 4 – 281 regions
```

----------------

Coordinates of clusters' centroids with respect to principal components of Factorial K-means

```
              Component 1   Component 2
[Cluster 1]   -0.6228974    1.700024531
[Cluster 2]    0.3154438    0.006924866
[Cluster 3]   -0.2611542   -0.991203260
[Cluster 4]    1.4111348   -2.444121954
```

-----------------

REGION WITH THE MINIMUM DISTANCE WITH RESPECT TO CENTROID OF CLUSTER 1

| | Region | V1 | V2 | output_cluster_for_euclidean_distance | RELATIVE CENTROID V1 COORDINATES | RELATIVE CENTROID V2 COORDINATES | SQUARED EUCLIDEAN DISTANCE WRT THE RELATIVE CENTROID |
|---|---|---|---|---|---|---|---|
| 1817 | ITF5 - Basilicata_2016 | -0.636459 | 1.703864 | 1 | -0.622897 | 1.700025 | 0.000199 |

-----------------

REGION WITH THE MINIMUM DISTANCE WITH RESPECT TO CENTROID OF CLUSTER 2

| | Region | V1 | V2 | output_cluster_for_euclidean_distance | RELATIVE CENTROID V1 COORDINATES | RELATIVE CENTROID V2 COORDINATES | SQUARED EUCLIDEAN DISTANCE WRT THE RELATIVE CENTROID | EUCLIDEAN DISTANCE CALCULATION WRT THE RELATIVE CENTROID |
|---|---|---|---|---|---|---|---|---|
| 754 | DE27 - Schwaben_2020 | 0.312302 | -0.03747 | 2 | 0.315444 | 0.006925 | 0.001981 | 0.044506 |

-----------------

## REGION WITH THE MINIMUM DISTANCE WITH RESPECT TO CENTROID OF CLUSTER 3

| | Region | V1 | V2 | output_cluster_for_euclidean_distance | RELATIVE CENTROID V1 COORDINATES | RELATIVE CENTROID V2 COORDINATES | SQUARED EUCLIDEAN DISTANCE WRT THE RELATIVE CENTROID | EUCLIDEAN DISTANCE CALCULATION WRT THE RELATIVE CENTROID |
|---|---|---|---|---|---|---|---|---|
| 1846 | NO07 - Nord-Norge_2016 | -0.280986 | -1.06179 | 3 | -0.261154 | -0.991203 | 0.005376 | 0.07332 |

----------------

## REGION WITH THE MINIMUM DISTANCE WITH RESPECT TO CENTROID OF CLUSTER 4

| | Region | V1 | V2 | output_cluster_for_euclidean_distance | RELATIVE CENTROID V1 COORDINATES | RELATIVE CENTROID V2 COORDINATES | SQUARED EUCLIDEAN DISTANCE WRT THE RELATIVE CENTROID | EUCLIDEAN DISTANCE CALCULATION WRT THE RELATIVE CENTROID |
|---|---|---|---|---|---|---|---|---|
| 38 | DE3 - Berlin_2023 | 1.629911 | -1.935844 | 4 | 1.411135 | -2.444122 | 0.306209 | 0.553362 |

----------------